\documentclass[twocolumn,eqsecnum,showpacs,showkeys,amsmath,amssymb,aps]{revtex4}

\usepackage{epsfig}
\usepackage{graphicx}
\usepackage{bm}

\makeatletter
\newcommand\erfc{\mathop{\operator@font erfc}\nolimits}
\def\slashchar#1{\setbox0=\hbox{$#1$}
   \dimen0=\wd0 \setbox1=\hbox{/} \dimen1=\wd1
   \ifdim\dimen0>\dimen1 \rlap{\hbox to \dimen0{\hfil/\hfil}} #1
   \else  \rlap{\hbox to \dimen1{\hfil$#1$\hfil}} / \fi}

\makeatother

\begin{document}
 
\title{Solution of the Kwieci\'nski evolution equations for 
unintegrated parton distributions using the Mellin transform\footnote{Dedicated to the
memory of Jan Kwieci\'nski}}

\author{Enrique Ruiz Arriola}
\email{earriola@ugr.es}
\affiliation{Departamento de F\'{\i}sica Moderna, Universidad de
Granada, E-18071 Granada, Spain}
\author{Wojciech Broniowski} 
\email{Wojciech.Broniowski@ifj.edu.pl} 
\affiliation{The H. Niewodnicza\'nski Institute of Nuclear Physics,
Polish Academy of Sciences, PL-31342 Krak\'ow, Poland}
\date{March 31, 2004}

\begin{abstract}
The Kwieci\'nski equations for the QCD evolution of the unintegrated
parton distributions in the transverse-coordinate space ($b$) are
analyzed with the help of the Mellin-transform method. The
equations are solved numerically in the general case, as well as in a
small-$b$ expansion which converges fast for $b \Lambda_{\rm QCD}$
sufficiently small. We also discuss the asymptotic limit of large $bQ$
and show that the distributions generated by the evolution decrease
with $b$ according to a power law. Numerical results are presented for the pion
distributions with a simple valence-like initial condition at the low
scale, following from chiral large-$N_c$ quark models. We use two models: the
Spectral Quark Model and the Nambu--Jona-Lasinio model. Formal aspects
of the equations, such as the analytic form of the $b$-dependent
anomalous dimensions, their analytic structure, as well as the limits
of unintegrated parton densities at $x \to 0$, $x \to 1$, and at large $b$, are
discussed in detail. The effect of spreading of the transverse
momentum with the increasing scale is confirmed, with $\langle
k_\perp^2 \rangle$ growing asymptotically as $Q^2
\alpha(Q^2)$. Approximate formulas for $\langle k_\perp^2 \rangle$ for
each parton species is given, which may be used in practical
applications.
\end{abstract}

 

\maketitle 
 
\section{Introduction\label{intro}}

The {\em unintegrated} parton distributions (UPD's) have been considered in 
numerous works on applications of 
perturbative QCD \cite{DDT,naka,sterman,kimb1,KMKS1,kimb2,MR,KHMR,WMR,MIU,collins,AS1,MarRys,JK1,JK2,JK3}. 
These distributions are generalizations of the 
usual integrated parton distributions (PD), and in some sense 
more basic objects, as PD's are obtained from UPD's by integration over the
transverse momentum of the parton. The notion of UPD's relies on the 
$k_\perp$-factorization, and, in the spirit of the 
CCFM equations \cite{CCFM1,CCFM2,CCFM3,CCFM4}, introduces two
scales: the probing scale, $Q$, and the transverse-momentum scale, $k_\perp$. 
Recently, the UPD gained substantial attention, since they enter many
exclusive physical processes, such as the production of the Higgs boson \cite{Jung,Gawron}, the 
$W$ boson \cite{JKAS}, 
heavy flavors \cite{Jung,Hansson,Motyka,Zotov,LipZot}, the
jet production \cite{Jung,Watt}, particle production \cite{AS}, or 
hadron production in relativistic heavy-ion collisions \cite{Biro,Levai}. The unintegrated distributions 
were also used in studies of the longitudinal \cite{Kotl}, charmed \cite{Kot}, and spin 
\cite{spin1,spin2} structure functions
of the nucleon, as well as analyzed in the dipole picture of QCD \cite{Shoshi}.

The UPD's, similarly to other entities in QCD, undergo evolution with the change of the probing scale $Q$. 
The philosophy adopted here is similar to the case of the integrated PD. At some initial scale $Q_0$
we need to know the non-perturbative 
quantity from measurements, models, or lattice calculations, and 
then we can evolve it to a different scale $Q$ with the help of suitable QCD evolution
equations. In the case of integrated PD we need to assume the dependence of PD on the 
Bjorken $x$ variable at the initial scale 
$Q_0$. For the UPD we need to know in addition the dependence on 
the transverse momentum, $k_\perp$, or, equivalently, the transverse coordinate $b$, which 
is the variable Fourier-conjugated to $k_\perp$. 
Knowing this, we compute, with no extra physical input apart for the assumptions
entering the QCD evolution, the unintegrated distribution at the final scale $Q$.
 
Important physics questions may be answered. In particular, the UPD 
evolve in such a way that the average transverse-momentum increases 
in a specified way with the scale \cite{JK1,JK2,JK3,Jung}. 
This spreading can be studied quantitatively within the approach. This constrains the freedom 
in phenomenological analyses of processes involving the UPD's. 

In his studies of the problem, Kwieci\'nski started from the CCFM formalism 
\cite{CCFM1,CCFM2,CCFM3,CCFM4},  which explicitly involves two separate scales: the probing scale, $Q$, and 
the transverse momentum of the parton, $k_\perp$. The CCFM equations were subsequently extended to include the quarks, 
as well as reduced to the single-loop approximation. In addition, the non-Sudakov
form factor was dropped. Kwieci\'nski found that 
in this approximation the 
equations diagonalize in the space Fourier-conjugated to the transverse 
momentum, where they assume a particularly simple and elegant form 
in a close resemblance of the DGLAP \cite{DGLAP1,DGLAP2,DGLAP3,DGLAP4} equations for the integrated PD. 
The only, but most important, difference is the appearance of the Bessel function $J_0(Q b)$ 
in the evolution kernel. 
Thus, the evolution depends on the transverse coordinate $b$.
In the original work of Ref.~\cite{JK3}
these equations were called ``the CCFM equations for the UPD in the transverse-coordinate
space in the single-loop approximation''. Due to numerous steps leading away from the original CCFM, 
we find it more appropriate to call the equations of Refs.~\cite{JK3} {\em the Kwieci\'nski 
equations for the UPD evolution}.  
Since in the case of integrated PD these equations reduce to the usual LO DGLAP,
the range of applicability of the Kwieci\'nski equations is not larger as for the LO DGLAP, with 
not too small and not too large values of the Bjorken $x$ variable.  

Formally, the Kwieci\'nski equations are integro-differential equations with the kernel depending on 
the transverse coordinate $b$  ({\em cf.} Sec. \ref{sec:keq}). 
As such, they are not trivial to solve 
numerically. The method of the original works \cite{JK1,JK2,JK3} involved the Chebyshev interpolation
in the $x$ and $Q$ spaces for each value of $b$. In this paper we offer an alternative method,
based on the evolution of the $x$-moments of UPD. In the moment (Mellin) space, the 
evolution equations become a set of ordinary differential equations, which can be solved 
numerically in a very efficient way (see Sec.~\ref{sec:keqmel}). We derive analytic expressions for the $b$-dependent
anomalous dimensions, which can be written in terms of hypergeometric 
functions. Then, the inverse Mellin transform to the original $x$-space is performed 
via numerical integration with oscillatory functions. We show that the procedure 
is fast and stable, providing a useful numerical tool for evolving the UPD.  

The Mellin-transform method allows us to carry  
analytic considerations, such as studies of certain limits of the
equations, specifically the cases of low and and large $b$, and $x$ approaching the endpoints.
We pursue these considerations, which can be done since the form of the $b$-dependent anomalous dimensions is 
analytic. 

In addition to developing a different numerical method
(Sec.~\ref{sec:keqmel}), our study differs from Ref.~\cite{JK3} in two
physical aspects. Firstly, rather than guessing the initial shape in
$b$, we use the results of low-energy chiral quark models
(Sec.~\ref{sec:init}).  We consider two models: the recently proposed
Spectral Quark Model of Ref.~\cite{SQM01,SQM,SQM03} and the
Nambu--Jona-Lasinio model with the Pauli Villars regularization
\cite{NJL}.  These models were used before to describe the
integrated PD \cite{DavArr1,DavArr2,Weigel,ERAZak}, and were shown to
do a surprisingly good job, in particular for the valence distribution
in the pion. They were also used to describe successfully other aspect
of high-energy processes, such as the pion distribution amplitude
\cite{PDA} and generalized (off-forward) PD of the pion
\cite{GPD,vento}.  The models give the initial condition at the model
scale $Q_0$ in a particularly simple, factorized form. The valence
quarks are distributed uniformly in $x$, while the gluons and the sea
quarks vanish. The $b$-dependence is a simple, analytic function with
exponential fall-off at large $b$.  We stress that the $b$ dependence
is a prediction of the model, rather than a mere guess, as is
frequently made.  Secondly, our implementation of the evolution
switches from three to four flavors above the charm-production
threshold, customarily taken at $Q^2= 4~{\rm GeV}^2$. Our numerical
results are presented in Sec.~\ref{sec:num}, where we show the
dependence of the UPD's on $x$ and $b$.

Since the analytic forms of the anomalous dimensions involve generalized 
hypergeometric functions, 
which may be cumbersome to program, we have developed a low-$b$ expansion (Sec.~\ref{sec:lowb}), 
which makes the calculations simpler when $b$ is not too large. 
The expansion is in powers of $b \Lambda_{\rm QCD}$, and is fast and stable.
For the opposite problem, where $Q b$ is large, we have obtained asymptotic expansions (Sec.~\ref{largeb}),
which allow to deal numerically with the generalized hypergeometric functions. 
  
Our method of solving the equations in the Mellin space carries additional bonuses. 
In particular, it allows us for analytic considerations in the investigation of formal limits
at $x \to 1$ (Sec.~\ref{sec:x1}) and $x \to 0$ (Sec.~\ref{sec:x0}).  
At $x \to 1$ we show that the $b$-dependent non-singlet 
distribution approaches very fast the 
integrated non-singlet distrtibution. At $x \to 0$  we find 
generalizations of the double-leading-logarithm (DLLA) 
formulas of Ref.~\cite{DLLA}. Finally, we examine the large-$b$ behavior, where we show
that the evolution-generated UPD's from the Kwieci\'nski equations 
exhibit power-law behavior at large $b$. The fall-off is much faster for the gluons than for the 
quarks. 

Widening in the transverse momentum of all partonic distributions is confirmed. 
We show that $\langle k_\perp^2 \rangle$ grows with the probing scale as $Q^2 \alpha(Q^2)$. 
We write an approximating formula for the width for each partonic species, which may be useful
in practical applications with the pion (Sec.~\ref{sec:kt}). 
The widening effect becomes stronger and stronger as $Q$ increases or $x$ decreases, and it is 
bigger
for the gluons than for the non-singlet and singlet quarks (see Sec.~\ref{sec:kt}).

The numerical method of this paper, which is easy to program and numerically fast and efficient, can be used for
other initial conditions as well, for instance for the GRS \cite{GRS} parameterization of the pion 
or the GRV parameterization \cite{GRV}, of the nucleon, supplied with a profile in the transverse coordinate, 
as originally studied in Ref.~\cite{JK3}. The only difference is in the form of the initial Mellin moments,
which acquire the dependence on $b$. General prediction of the method in formal limits are listed in
Sec.~\ref{sec:general}.

Appendices contain many technical details, such as the perturbative QCD parameters and splitting
functions (App.~\ref{pQCD}), the analytic form of 
the $b$-dependent anomalous dimensions which 
enter the evolution in the Mellin space (App.~\ref{anomdim}), their low-$b$ 
(App.~\ref{app:lowb}) and high-$b$ expansion (App.~\ref{app:largeb}), as well as the pole-residue 
expansion (App.~\ref{sec:polres}). The latter is useful in analytic considerations near $x=0$.

\section{The Kwieci\'nski equations\label{sec:keq}}

In his studies of the UPD's, Kwieci\'nski \cite{JK1,JK2,JK3,Gawron} started from the CCFM formalism 
\cite{CCFM1,CCFM2,CCFM3,CCFM4} explicitly involving two separate scales: the probing scale, $Q$, and 
the transverse momentum of the parton, $k_\perp$. Then, the original CCFM equations were supplemented with the quarks, 
as well as reduced to the single-loop approximation. The latter approximation replaces the angular ordering of 
the emitted gluons with the ordering of their transverse momenta. 
In addition, the non-Sudakov form factor was dropped. 
Kwieci\'nski realized that the evolution equations for the 
UPD's acquire a particularly simple form in the transverse-coordinate space, $b$, 
conjugated to the transverse momentum $k_\perp$.
For each distribution one introduces
\begin{equation}
f_{j}(x,b,Q)\,=\,\int_0^{\infty}2 \pi dk_\perp \,k_\perp\, J_0(b k_\perp)\, f_{j}(x,k_\perp,Q),
\label{fb2}
\end{equation}
where $j=NS$ (non-singlet quarks), $S$ (singlet quarks), or $G$ (gluons),  and $J_0$ is the Bessel function. 
In order to avoid confusion, we stress 
that the transverse coordinate $b$, conjugated to the parton's transverse momentum,
is not the impact parameter, appearing in the analysis of the generalized PD. The latter 
quantity is conjugated the the transverse momentum transfer in off-forward 
scattering processes. 

At $b=0$ the functions $f_{j}$ are related to the integrated parton distributions, $p_j(x,Q)$,
as follows: 
\begin{equation}
f_{j}(x,0,Q)=\frac{x}{2} p_j(x,Q). \label{deff}
\end{equation}
More explicitly, for the case of the pion studied in this paper 
(we take $\pi^+$ for definiteness) we have
\begin{eqnarray}
p_{\rm NS}&=&\bar u - u + d -\bar d, \nonumber \\
p_{S}&=&\bar u + u + d +\bar d +\bar s + s + ... , \label{partpi}\\
p_{\rm sea}&\equiv& p_S - p_{\rm NS} = 2 \bar d + 2 u +\bar s + s + ... , \nonumber \\
p_G &=& g, \nonumber
\end{eqnarray}
where $\dots$ stand for higher flavors.

The Kwieci\'nski equations read \cite{JK3}:
\begin{widetext}
\begin{eqnarray}
Q^2{\partial f_{\rm NS}(x,b,Q)\over \partial Q^2} &=&
{\alpha_s(Q^2)\over 2\pi}  \int_0^1dz  \,P_{qq}(z)
\bigg [ \Theta(z-x)\,J_0((1-z)Qb)\, f_{\rm NS}\left({x\over z},b,Q\right)
- f_{\rm NS}(x,b,Q) \bigg ] \nonumber \\ 
Q^2{\partial f_{S}(x,b,Q)\over \partial Q^2}
&=&
{\alpha_s(Q^2)\over 2\pi} \int_0^1 dz
\bigg\{\Theta(z-x)\,J_0((1-z)Qb)\bigg[P_{qq}(z)\, f_{S}\left({x\over z},b,Q\right)
+ P_{qG}(z)\, f_{G}\left({x\over z},b,Q\right)\bigg]
\nonumber \\ &-& [zP_{qq}(z)+zP_{Gq}(z)]\, f_{S}(x,b,Q)\bigg\}
\nonumber \\
Q^2{ \partial  f_{G}(x,b,Q)\over \partial Q^2}&=&
{\alpha_s(Q^2)\over 2\pi} \int_0^1 dz
\bigg\{\Theta(z-x)\,J_0((1-z)Qb)\bigg[P_{Gq}(z)\, f_{S}\left({x\over z},b,Q\right)
+ P_{GG}(z)\, f_{G}\left({x\over z},b,Q\right)\bigg]
\nonumber \\ 
&-& [zP_{GG}(z)+zP_{qG}(z)]\, f_{G}(x,b,Q)\bigg\}
\label{Keq}
\end{eqnarray}
\end{widetext}
The splitting functions $P_{ab}(z)$ are listed in Eq.~(\ref{split}). 

Following Ref. \cite{JK3},  a factorized form 
of the distribution functions at the initial scale $Q_0$ is assumed, 
\begin{equation}
 f_j(x,b,Q_0)=F^{\rm NP}(b)\frac{x}{2}p_j(x,Q_0), 
\label{bcondf}
\end{equation}
with the profile function $F^{\rm NP}(b)$ taken to be universal for all species
of partons.  The factorization assumption (\ref{bcondf}) is technical and one can easily depart
from this limitation in numerical studies.  
We note, however, that the models of Ref. \cite{SQM,NJL},
studied in Sec.~\ref{sec:init}, do predict a factorized initial
condition of the form (\ref{bcondf}).  The input profile function, $F^{\rm NP}(b)$, is linked through the
Fourier-Bessel transform to the $k_\perp$ distribution at the scale
$Q_0$.  At $b=0$ the normalization is $F^{\rm NP}(0)=1$.  The profile function
factorizes from the evolution equations. This is clear, since 
any solution of Eq.~(\ref{Keq}) remains a solution when multiplied by an arbitrary function of $b$. 
Due to evolution, at higher scales $Q$ we have
\begin{eqnarray}
f_j(x,b,Q)=F^{\rm NP}(b) f_j^{\rm evol}(x,b,Q), \label{factor}
\end{eqnarray}
with $f_j^{\rm evol}(x,b,Q)$ satisfying equations {\em identical} to (\ref{Keq}).

We should stress again an important physical difference between $F^{\rm NP}(b)$ and $f_j^{\rm evol}(x,b,Q)$. 
While $F^{\rm NP}(b)$ originates entirely from low-energy, 
{\em non-perturbative} physics, $f_j^{\rm evol}(x,b,Q)$ 
is given by the {\em perturbative QCD evolution} with Eq.~(\ref{Keq}) from the initial condition
\begin{equation}
 f_j^{\rm evol}(x,b,Q_0)=\frac{x}{2}p_j(x,Q_0). 
\label{bcondfevol}
\end{equation}
Throughout this paper, except for Sec.~\ref{sec:init}, we focus on the perturbative evolution and the functions
$f_j^{\rm evol}(x,b,Q)$. The function $F^{\rm NP}(b)$ is referred to as {\em the initial profile} and 
$f_j^{\rm evol}(x,b,Q)$ as {\em the evolution-generated UPD}.


\section{The Kwieci\'nski equation in the Mellin space\label{sec:keqmel}}

We define the $x$-moments of an unintegrated parton distribution in the impact
parameter space as (we retain the same symbol for the function and its Mellin 
transform, hoping the distinction made by the argument prevents any confusion)  
\begin{equation}
f_j(n, b, Q) = \int_0^1 dx \,x^{n-1} f_j(x, b, Q). \label{momdef}
\end{equation}
In the Mellin space, the evolution
equations for the UPD are very simple, as they become diagonal both in $b$ and $n$. 
They involve $b$-dependent anomalous dimensions, equal to
\begin{eqnarray}
&& \gamma_{n,ab}( Q b ) = 
4 \int_0^1 dz \, \left[ z^n J_0 \left( (1-z) Q b \right) -1 \right] P_{ab} (z)  \nonumber \\
&& = \gamma_{n,ab}^{(0)} \label{gamdef} 
- 4 \int_0^1 dz \, z^n \left[ J_0 \left( (1-z) Q b \right) -1 \right] P_{ab} (z),  \nonumber \\
\label{gammadef}
\end{eqnarray}
where the values at $b=0$ are
\begin{eqnarray}
\gamma_{n,NS}^{(0)} &=& 
- 4 \int_0^1 dz (z^n -1) P_{qq} (z), \nonumber \\
\gamma_{n,qq}^{(0)} &=& 
- 4 \int_0^1 dz \left [ (z^n -z)P_{qq}-z P_{Gq} \right], \nonumber \\
\gamma_{n,qG}^{(0)} &=& - 4 \int_0^1 dz z^n P_{qG},  \\
\gamma_{n,Gq}^{(0)} &=& - 4 \int_0^1 dz z^n P_{Gq}, \nonumber \\
\gamma_{n,GG}^{(0)} &=& - 4 \int_0^1 dz \left [ (z^n-z) P_{GG} -z P_{qG} \right ]. \nonumber
\label{gamdef0}
\end{eqnarray} 
Their explicit form for various channels is listed in
Eq.~(\ref{anombNS},\ref{anomb}).  The fact that we can write the
analytic form of the integrals in Eq. (\ref{gammadef}) (see
App.~\ref{anomdim}) allows for efficient numerical calculations and
analytic considerations.

The integration of both sides of Eq.~(\ref{Keq}) with $\int_0^1 dx
x^{n-1}$ yields for the non-singlet case the equation
\begin{equation}
\frac{ d f_{\rm NS}(n, b, Q )}{ dQ^2 } = -\frac{\alpha_S(Q^2)}{8 \pi Q^2}
\gamma_{n,NS}(Q b ) f_{\rm NS}(n, b,Q). 
\label{eqNS}
\end{equation} 
The formal solution of Eq.~(\ref{eqNS}) can be readily obtained as
\begin{equation}
\frac{ f_{\rm NS}(n, b, Q)}{f_{\rm NS}(n, b, Q_0) } = {\rm exp} \left[
-\int_{Q_0^2}^{Q^2} \frac{d{Q'}^2 \alpha({Q'}^2)}{8\pi {Q'}^2}
\gamma_{\rm NS}(n, b, Q') \right].
\label{nschan}
\end{equation}
In the singlet channel we find the coupled set of equations, 
\begin{eqnarray}
&& \!\!\!\!\! \frac{ d f_{S}(n, b, Q )}{ dQ^2 } =
-\frac{\alpha_S(Q^2)}{8 \pi Q^2} \nonumber \\ && \times \left [
\gamma_{n,qq}(Q b ) f_{S}(n, b,Q) + \gamma_{n,qG}(Q b ) f_{G}(n, b,Q)
\right ], \nonumber \\ && \!\!\!\!\! \frac{ d f_{G}(n, b, Q )}{ dQ^2 }
= -\frac{\alpha_S(Q^2)}{8 \pi Q^2} \nonumber \\ && \times \left [
\gamma_{n,Gq}(Q b ) f_{S}(n, b,Q) + \gamma_{n,GG}(Q b ) f_{G}(n, b,Q)
\right ], \nonumber \\
\label{eqS}
\end{eqnarray} 
which has the formal solution
\begin{eqnarray}
&&\left ( \begin{array}{c} f_S(n,b,Q) \\f_G(n,b,Q) \end{array} \right ) = \label{eqSformal} \\
&& {\cal P} \exp \left [ {-\int_{Q_0^2}^{Q^2} \frac{dQ'^2 \alpha(Q'^2)}{8 \pi Q'^2} 
\Gamma_n(Qb)} \right] \left ( \begin{array}{c} f_S(n,b,Q_0) \\f_G(n,b,Q_0) \end{array} \right ), \nonumber \\
&& \Gamma_n(Qb)=\left ( \begin{array}{cc} \gamma_{n,qq}(Qb) & \gamma_{n,qG}(Qb) \\ 
\gamma_{n,Gq}(Qb) & \gamma_{n,GG}(Qb) \end{array} \right ). \nonumber
\end{eqnarray}
The symbol ${\cal P}$ indicates that powers of $\Gamma_n$ are ordered along the integration path.
The qualitative difference between Eq.~(\ref{eqSformal}) and the LO DGLAP equations is the fact that 
$\Gamma_n$ depends on the evolution variable $Q$. This makes the singlet sector more difficult
to analyze analytically. Equations~(\ref{eqS}) can be solved numerically for any value of $n$ real or
complex~\cite{ERANLO}. For the case of integrated PD ($b=0$)
Eq.~(\ref{eqNS}) and (\ref{eqS}) reduce to the well-known LO DGLAP
equation in the Mellin space.

The corresponding UPD in the $x$ space can be reconstructed using
the inverse Mellin transform, 
\begin{equation}
f_j(x, b, Q) = \int_{n_0-i\infty}^{n_0+i\infty}\frac{{\rm d} n}{2\pi{i}}
x^{-n} f_j(n, b, Q), \label{invmel}
\end{equation}
where $n_0$ has to be chosen in such a way as to leave all the
singularities on the left-hand-side of the contour. It turns out that
for $ b \neq 0 $ the analytic structure of the $b-$dependent anomalous
dimension remains the same as for the $b=0$ case. This can be inferred
directly by studying the analytic structure of the formulas
(\ref{anombNS},\ref{anomb}) or the pole-residue expansion of
Eq.~(\ref{eq:pole_res}). Thus, we have the result that that
$\gamma_{n,NS}(Q b)$, $\gamma_{n,qq}(Q b)$, and $\gamma_{n,qG}(Q b)$
have poles at $n=-1, -2, -3, ...$, while $\gamma_{n,GG}(Q b)$ and
$\gamma_{n,Gq}(Q b)$ have poles at $n=0, -1, -2, ...$. Parameterizing
the contour in Eq.~(\ref{invmel}) as $n=n_0+i t$, we arrive at the
inversion formula
\begin{eqnarray}
f_j(x, b, Q) \!\!&=& \!\! x^{-n_0} \!\! \int_0^{\infty} \!\! \frac{{\rm d} t}{\pi}
 \left ( \cos(t \log x) {\rm Re}[f_j(n_0+i t, b, Q)]\nonumber \right . \\ 
&+& \left . \sin(t \log x) {\rm Im}[f_j(n_0+i t, b, Q)] \right ).  \label{invmel2}
\end{eqnarray}
For the non-singlet case we take $n_0=0$, while for the singlet case
$n_0=1$. As an additional check we have also verified that a bended
integration path $ n = c + r e^{i\pi /4} $ with $0 \le r < \infty$, as
used in \cite{ERANLO}, also works nicely. 

\section{The initial condition for the UPD of the pion\label{sec:init}} 

As an illustration of our method, we consider the UPD's of the pion with
the initial condition at $Q_0$ provided by two large-$N_c$ low-energy chiral quark
models. The original work of Ref.~\cite{JK3} tested the GRS
parameterization for the pion \cite{GRS} and the GRV parameterization
for the nucleon \cite{GRV}, supplied with a Gaussian profile $F^{\rm NP}(b)$.
The considered models generate the functions $F^{\rm NP}(b)$ as genuine model
predictions of low-energy non-perturbative physics, with no freedom
involved.  Our first model is the recently proposed Spectral
Quark Model (SQM)\cite{SQM01,SQM,SQM03}, and the second one is the
popular Nambu--Jona Lasinio model (NJL) with the Pauli-Villars
regularization \cite{NJL}, treated for simplicity in the strict chiral
limit.  

Firstly, since the chiral quark models have no gluon degrees of freedom, 
we have at the model scale 
\begin{eqnarray}
g(x,Q_0)=0 ,  \label{noglue}
\end{eqnarray}
or 
\begin{eqnarray}
f_{G}^{\rm evol}(x,b,Q_0)= 0 . \label{noglue2}
\end{eqnarray}
For the integrated valence quark PD both models predict, in the chiral
limit and at the model scale $Q_0$, that
\begin{eqnarray}
q(x,Q_0)=\theta(x) \theta(1-x), \label{thetas}
\end{eqnarray}
{\em i.e.} a constant value with the proper normalization and support.
Thus the corresponding $f^{\rm evol}$ functions of Eq.~(\ref{deff}) are linear in $x$, 
\begin{eqnarray}
f_{NS,S}^{\rm evol}(x,b,Q_0)=x \theta(x) \theta(1-x), \label{initx}
\end{eqnarray}
The scale of the model, as found from the momentum sum-rule, is rather
low: $Q_0=313$~MeV.  Although this is admittedly a very low scale, one
may hope that the typical expansion parameter, $\alpha(Q_0^2)/(2 \pi)
\sim 0.3$, is low enough to make the perturbation theory sensible.
This claim gains support from the next-too-leading analysis of the
integrated PD \cite{DavArr1}, where the corrections are found to be
small.

The QCD evolution is crucial for the phenomenological success of the
considered low-energy chiral quark models.  In
Refs. \cite{DavArr1,DavArr2,Weigel} it has been found that the
non-singlet distribution, when evolved to the scale of 2~GeV, agrees
very well with the SMRS parameterization of the pion data
\cite{SMRS92}, while in \cite{WBDur} it has been very favorably
compared to the old E615 data at 4~GeV \cite{e615} (see the discussion in
Sec.~\ref{sec:num} and Fig.~\ref{fig:e615}).
  
In the SQM, the valence UPD of the pion at the scale $Q_0$ is
\cite{SQM}
\begin{eqnarray}
q(x,k_\perp,Q_0)&=&\overline{q}(1-x,k_\perp) \label{qqmod} \\
&=&\frac{6M_V^3}{\pi(k_\perp^2+M_V^2/4)^{5/2}}
 \theta(x)\theta(1-x), \nonumber
\end{eqnarray}
where $M_V=770~{\rm MeV}$ is the mass of the $\rho$ meson.
Passing to the impact-parameter space with the Fourier-Bessel transform yields
\begin{eqnarray}
q(x,b,Q_0)& \equiv & 2 \pi \int_0^\infty k_\perp d k_\perp q (x,k_\perp) J_0(k_\perp b) \label{sqmb} \\
&=& \left (1+\frac{b M_V}{2} \right )
 \exp\left ( - \frac{M_V b}{2} \right ) 
 \theta(x)\theta(1-x). \nonumber
\end{eqnarray}
The expansion at small $b$ gives
\begin{eqnarray}
 q(x,b,Q_0)&=& \left( 1-\frac{M_V^2 b^2}{8}+\frac{M_V^3 b^3}{24}+\dots
\right) \nonumber \\ && \times \theta(x)\theta(1-x), 
\label{Psibexp}
\end{eqnarray}
and average transverse momentum squared is equal to  
\begin{eqnarray}
\langle k_\perp^2 \rangle_{\rm NP}^{\rm SQM} &\equiv& \frac{\int d^2 k_\perp \, k_\perp^2 q(x,k_\perp)}
{\int d^2 k_\perp q(x,k_\perp)}=-\frac{4}{q(x,b)}\left.\frac{d q(x,b)}{db^2}\right|_{b=0} \nonumber \\
&=& \frac{M_V^2}{2}, \label{kperp2}
\end{eqnarray}
which numerically gives $\langle k_\perp^2 \rangle_{\rm NP}^{\rm SQM}=(544~{\rm MeV})^2$ (all at the model working scale
$Q_0$). The subscript {\em NP} reminds 
us that the quantity comes entirely from the non-perturbative physics, 
entering profile function $F^{\rm NP}(b)$ (see the discussion at the end of Sec.~\ref{sec:keq}).

In the NJL model with the PV regularization \cite{NJL} the analogous
formulas read~\cite{ERAZak} 
\begin{eqnarray}
&& q(x,k_\perp,Q_0)=\overline{q}(1-x,k_\perp) \label{qqNJL} \\
&& =\frac{\Lambda^4 M^2 N_c}
  {4 f_\pi^2 \pi^3 \left( k_\perp^2 + M^2 \right) \left( k_\perp^2 + \Lambda^2 + M^2 \right)^2} 
\theta(x)\theta(1-x), \nonumber
\end{eqnarray}
\begin{eqnarray}
q(x,b,Q_0)&=& \frac{M^2 N_c}{4 f_\pi^2 \pi^2} \left( 2 K_0(b M) - \label{qqNJLb}
        2 K_0(b \sqrt{\Lambda^2 + M^2}) \nonumber \right . \\ &-& \left . 
        \frac{b \Lambda^2 K_1(b \sqrt{\Lambda^2 + M^2})}
{\sqrt{\Lambda^2 + M^2}} \right ) \theta(x)\theta(1-x), \nonumber \\
\end{eqnarray}
where the pion decay constant is given by
\begin{eqnarray}
f_\pi^2=\frac{ M^2 N_c \left( \Lambda^2 + 
        \left( \Lambda^2 + M^2 \right)  \log \frac{\Lambda^2 + M^2}{M^2}  
\right)}{4 \pi^2 \left( \Lambda^2 + M^2 \right) }. \label{fpi2}
\end{eqnarray}
The parameters of the model are adjusted in such a way that $f_\pi=93$~MeV, namely 
$M=280$~MeV and $\Lambda=871$~MeV.
The expansion at small $b$ yields
\begin{eqnarray}
&& q(x,b,Q_0) = \\ && \left( 1- \frac{M^2 N_c \left( \Lambda^2 - 
M^2 \log \frac{\Lambda^2 + M^2}{M^2}  \right) b^2
      }{16 \pi^2 f_\pi^2 } + \dots \right) \nonumber \\
&& \times 
\theta(x)\theta(1-x), \nonumber
\label{PsibexpNJL}
\end{eqnarray}
and average transverse momentum squared is equal to  
\begin{eqnarray}
\langle k_\perp^2 \rangle_{\rm NP}^{\rm NJL} = \frac{M^2 N_c \left( \Lambda^2  - 
M^2 \log \frac{\Lambda^2 + M^2}{M^2} \right) }
  {4 \pi^2 f_\pi^2}, \label{kperp2NJL}
\end{eqnarray}
which numerically gives $\langle k_\perp^2 \rangle_{\rm NP}^{NJL}=(626~{\rm MeV})^2$ (at the scale
$Q_0$), which is similar to the number from the SQM.

\begin{figure}[tb]
\begin{center}
\epsfig{figure=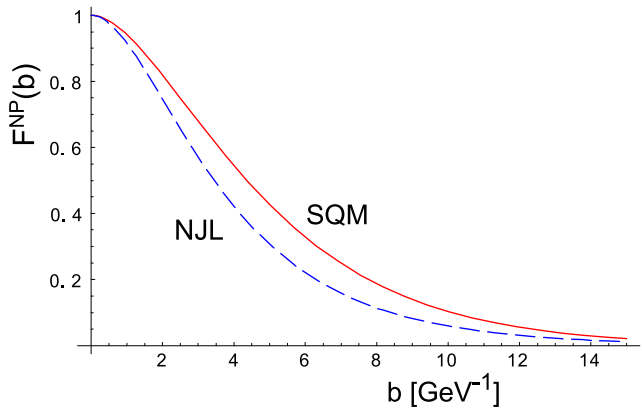,width=8.5cm}
\end{center}
\caption{
The initial profile functions $F_{\rm NP}$ for the SQM and NJL models, plotted as functions of the transverse coordinate $b$.
The fall-off is exponential, according to Eq.~(\ref{qqSQMb}) and (\ref{qqNJLb}).}
\label{fig:F}
\end{figure}

Finally, in the notation of Eq.~(\ref{factor}) we can write that the initial profile function is 
\begin{eqnarray}
F^{\rm NP}_{\rm SQM}(b)&=&\left (1+\frac{b M_V}{2} \right ) \exp\left ( - \frac{M_V b}{2} \right ), \label{qqSQMb}\\
F^{\rm NP}_{\rm NJL}(b)&=&\frac{M^2 N_c}{4 f_\pi^2 \pi^2} \left( 2 K_0(b M) - \label{qqNJLb2}
        2 K_0(b \sqrt{\Lambda^2 + M^2}) \nonumber \right . \\ &-& \left . 
        \frac{b \Lambda^2 K_1(b \sqrt{\Lambda^2 + M^2})}{\sqrt{\Lambda^2 + M^2}} \right ).
\end{eqnarray}
Both initial profile functions are displayed in Fig.~\ref{fig:F}.
Note that the profiles, although have a $b^2$ correction at small $b$,
are not Gaussian, and at large $b$ display an exponential fall-off.

As discussed in Sec. \ref{sec:keq}, the form of $F^{\rm NP}(b)$ factorizes from the evolution.
In both models there is no dependence of UPD on $x$ at the initial scale $Q_0$. As a results, we get the 
following set of initial moments:
\begin{eqnarray}
f_{\rm NS}^{\rm evol}(n,b,Q_0)&=&\frac{1}{n+1}, \nonumber \\
f_{S}^{\rm evol}(n,b,Q_0)&=&\frac{1}{n+1}, \nonumber \\
f_{G}^{\rm evol}(n,b,Q_0)&=&0. \label{initmom}
\end{eqnarray} 

We remark that away from the chiral limit the separability of the
dependence on $x$ and $b$ no longer holds.  In this case the initial
conditions for the evolution are more complicated (they depend on $b$), but the analysis
can be easily generalized to account for this case as well. 

\section{Numerical results\label{sec:num}}

\begin{figure}[tb]
\begin{center}
\epsfig{figure=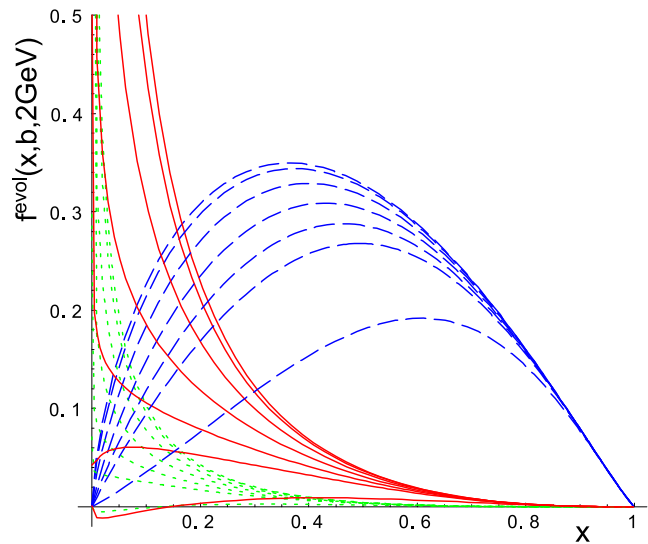,width=8.6cm}
\vspace{-10mm}
\end{center}
\caption{The evolution-generated UPD's for the pion for various values of the transverse
coordinate (from top to bottom $b=0$, $1$, $2$, $3$, $4$, $5$ and
$10~{\rm GeV}^{-1}$), plotted as functions of the Bjorken $x$.  The evolution is
made with the initial condition (\ref{initx}) at $Q_0=313$~MeV up to
$Q=2~{\rm GeV}$.  Solid lines -- gluons, dashed lines -- valence
quarks, dotted lines -- sea quarks.}
\label{Q2}
\end{figure}

In Fig. \ref{Q2} we present the results of our numerical calculation
with the method using the Mellin transform. The initial conditions are
for the pion in the chiral limit (\ref{initmom}), holding at
$Q_0=313$~MeV, and the evolution is carried up to the scale of
$Q=2$~GeV. The differential equations for the moments,
(\ref{eqNS},\ref{eqS}), are solved numerically for complex $n-$values
along the Mellin contour, and subsequently the inverse Mellin
transform (\ref{invmel2}) is carried out.  The figure contains three
families of curves, solid for the gluon, dashed for the valence
quarks, and dotted for the sea quarks. In each family $b$ assumes the
values 0, 1, 2, 3, 4, 5, and 10. Naturally, increasing $b$ results in
a decrease of the distribution, with the effect strongest at low
$x$. At $x$ close to 1 this effect disappears, which is explained in
Sec.~\ref{sec:x1}. We also note that at high values of $b$ the
distributions for the gluons become negative, reflecting the change of
sign of the Bessel function in the evolution kernel of
Eq.~(\ref{Keq}).  As already discussed in Ref.~\cite{JK3}, this poses
no immediate physical problems, as the distributions in $k_\perp$
remain positive as primary objects, and so are the physical cross sections.

The results of Fig.~\ref{Q2} are consistent with the findings 
of Ref. \cite{JK3}, where a different numerical method was used, as
well as a different initial condition tested. 

\begin{figure}[tb]
\begin{center}
\epsfig{figure=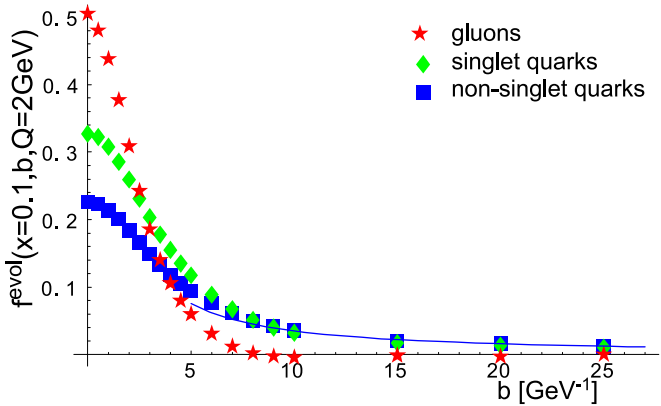,width=8.6cm}
\vspace{-10mm}
\end{center}
\caption{Evolution-generated UPD's for the pion for $Q=2$~GeV
and $x=0.1$, plotted as functions of $b$. The evolution is made with
the initial condition (\ref{initx}) at $Q_0=313$~MeV.  The numerical
results are represented by squares for the non-singlet quarks, diamond for
the singlet quarks, and stars for the gluons, while the solid line shows the
asymptotic formula (\ref{fnsas3} for the case of non-singlet quarks. We note the 
much faster fall-of for the gluons than for the quarks, as expected from Eq.~(\ref{powG}).
The quarks exhibit a long-range tail, according to the power-law formula (\ref{fnsas3}). 
As $Q$ is increased or $x$ decreased, the distributions in $b$ become narrower, leading to 
spreading in $k_\perp$. }
\label{fig:b}
\end{figure}

In Fig.~\ref{fig:b} we show the dependence of the evolution-generated UPD's on $b$ at
$Q=2$~GeV and $x=0.1$.  The
results are represented by squares for the non-singlet quarks, diamonds for
the singlet quarks, and stars for the gluons. We note the 
much faster fall-off with $b$ for the gluons than for the quarks, as expected from Eq.~(\ref{powG}).
The quarks exhibit a long-range tail, according to the power-law formula (\ref{fnsas3}). 
The solid line
shows the asymptotic form for the case of non-singlet quarks from Eq.~(\ref{fnsas3},\ref{fnsas4}) (see
Sec.~\ref{largeb}), which becomes accurate for $b \ge 10~{\rm GeV}^{-1}$. 
As $Q$ is increased further, or $x$ decreased, the distributions in $b$ become narrower, leading to 
larger spreading in $k_\perp$. For more details and plots concerning other numerical results see Ref.~\cite{JK3}. 

\section{Low-$b$ expansion\label{sec:lowb}}

\begin{figure}[]
\begin{center}
\epsfig{figure=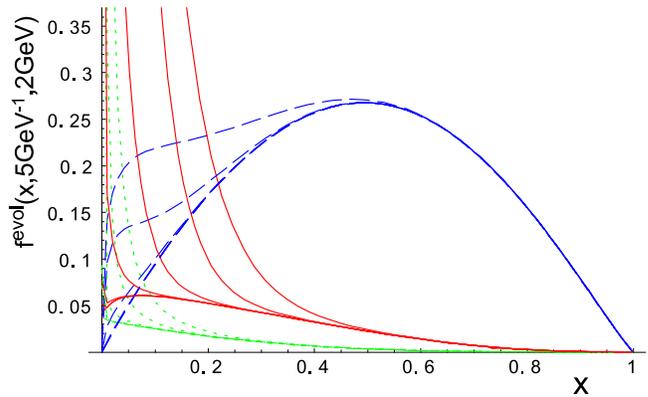,width=8.6cm}
\end{center}
\caption{The low-$b$ expansion for the evolution-generated UPD's of the pion
at $b=5~{\rm GeV}^{-1}$ and $Q=2~{\rm GeV}^2$, plotted as a function of the 
Bjorken $x$. 
Solid lines -- gluons, dashed lines -- valence quarks, dotted lines -- sea quarks.
For each kind of parton the curves from top to bottom correspond to 2, 4, 6, 8, 10, and 16
terms in the small $b$-expansion. The initial condition for the evolution is provided 
by Eq.~(\ref{initx}).}
\label{fig:convergence}
\end{figure}

Since the anomalous dimensions (\ref{anombNS},\ref{anomb}) involve 
generalized hypergeometric functions, they are not 
easy to use in numerical calculations. For that reason we consider the small-$b$ expansion, as well
as the asymptotic forms at large $b Q$, presented in the next section.
It is convenient to introduce the notation
\begin{eqnarray} 
r_k=r_k(Q_0^2,Q^2)=\int_{Q_0^2}^{Q^2} \frac{d{Q'}^2 \alpha({Q'}^2)}{8
\pi {Q'}^2} {Q'}^{2k}. \nonumber \\ \label{rk}
\end{eqnarray}
The explicit form of functions $r_k$ is given in Appendix \ref{pQCD}.
Next, we apply Eq.~(\ref{ge}) and find the following expansion in the non-singlet channel:
\begin{eqnarray}
&&\frac{ f(n, b, Q)}{f(n,b,Q_0) } = e^{\gamma_{n,NS}^{(0)} r_0} 
\exp \left[- C_F \sum_{k=1}^\infty \frac{(- b^2)^{k} 4^{1-k}}{k!^2} \right . \nonumber  \\
&& \left . \times \left[ {\sf B} (2k,n+1) + {\sf B}
(2k,n+3) \right] r_k \right] \label{fbf0},
\end{eqnarray}
where ${\sf B}$ is the Euler beta function. 
We note that although at the
level of the differential equation our expansion is formally in $Q b$,
the result (\ref{fbf0}), together with the fact that $r_k$ is proportional to 
$\Lambda_{\rm QCD}^{2k}$ ({\em cf.} (\ref{rkap})), show that the expansion 
parameter is actually $b \Lambda_{\rm QCD}$.
The rate of convergence of the
method can be deduced from Fig.~(\ref{fig:convergence}). As we can see, the
number of terms needed increases for increasing $b$ and decreasing
$x$. For $x > 0.01$ eight terms in the expansion seems more than sufficient.

\section{Large-$b$ expansion\label{largeb}}

Appendix \ref{app:largeb} contains asymptotic forms of the generalized hypergeometric
functions and of the $b$-dependent anomalous dimensions. These expressions hold
in the limit of large $Q b$ at $n$ kept fixed. The formulas are of 
great practical importance in the present study, since they are much 
simpler to implement in numerical calculations than the generalized hypergeometric 
functions appearing in Eq.~(\ref{anombNS},\ref{anomb}). Actually, Eq.~(\ref{asqq}-\ref{asGG}) 
should be used whenever $Q b$ is larger than about $10 |n|$. Since in practical problems the scale $Q$ 
may be as large as the mass of the $W$-boson, there is a frequent need to use the asymptotic 
expressions (\ref{asqq}-\ref{asGG}). Also, if the UPD's in the 
transverse momentum are needed, one has to carry back the Fourier transform from the $b$-space to
the $k_\perp$-space, which involves all values of $b$. 

We start with the non-singlet case of Eq.~(\ref{eqNS}).
With the help of Eq.~(\ref{asqq}), where at the leading order in $1/(Qb)$ we drop the oscillatory parts, 
we may write (\ref{nschan}) at large $Q b$: 
\begin{eqnarray}
\frac{f^{\rm evol}_{\rm NS}(n,b,Q)}{f^{\rm evol}_{\rm NS}(n,b,Q_0)}&=&e^{y_0+y_1 n}, \label{fasy}
\end{eqnarray}
where
\begin{eqnarray}
y_1&=& -4C_F \left ( \frac{2 \Lambda_{\rm QCD}^2}{b}-\frac{1}{b^3} \right ) r_{-3/2}(Q_0^2,Q^2) , \nonumber \\
y_0&=& -4C_F \left ( \frac{2 \Lambda_{\rm QCD}^2}{b} r_{-3/2}(Q_0^2,Q^2) 
+\frac{1}{\beta_0^{(3)}} \log \frac{Q}{Q_0} \right . \nonumber \\
&+& \left . 2 \left [ \log \frac{b \Lambda_{\rm QCD}}{2} +\gamma -\frac{3}{4} \right ] r_0(Q_0^2,Q^2)    \right ) ,
\nonumber \\
&& \hspace{4cm}  {\rm for} \;\; Q^2 < \mu_c^2, \label{y0y13}
\end{eqnarray}  
and 
\begin{eqnarray}
y_1&=& -4C_F \left ( \frac{2 \Lambda_{\rm QCD }^2}{b}r_{-3/2}(Q_0^2,\mu_c^2) \right . \nonumber \\
&+& \left . \frac{2 \Lambda_{4}^2}{b}r_{-3/2}(\mu_c^2,Q^2)-\frac{1}{b^3} r_{-3/2}(Q_0^2,Q^2) \right ) , \nonumber \\
y_0&=& -4C_F \left ( \frac{2 \Lambda_{\rm QCD}^2}{b} r_{-3/2}(Q_0^2,\mu_c^2) 
\right . \nonumber \\ &+& \left . \frac{2 \Lambda_{4}^2}{b} r_{-3/2}(\mu_c^2,Q^2)
+\frac{1}{\beta_0^{(3)}} \log \frac{\mu_c}{Q_0} + \frac{1}{\beta_0^{(4)}} \log \frac{Q}{\mu_c}  \right . \nonumber \\
&+& \left . 2 \left [ \log \frac{b \Lambda_{\rm QCD}}{2} \right ] r_0(Q_0^2,\mu_c^2) \right .  \nonumber \\ &+& \left . 
2 \left [ \log \frac{b \Lambda_{4}}{2} \right ] r_0(\mu_c^2,Q^2) 
 + \left [ \gamma -\frac{3}{4} \right ]   r_0(Q_0^2,Q^2)  \right ) ,
\nonumber \\
&&  \hspace{4cm}  {\rm for} \;\; Q^2 \ge \mu_c^2. \label{y0y14}
\end{eqnarray}  
Next, we take our initial condition (\ref{initmom}) and use the inverse Mellin transform 
\begin{eqnarray}
\int_C \frac{dn}{2\pi i} x^{-n} \frac{e^{y_0+y_1 n}}{n+1} = x e^{y_0-y_1} \Theta \left (y_1+\log \frac{1}{x} \right ).
\label{lapy}
\end{eqnarray}
The condition provided by the theta function means that the formula can be used for $x < \exp(y_1)$. For 
negative $y_1$ this means that the validity is limited for $x$ not too close to 1. This, however, has been 
already tacitly assumed, since the asymptotic expansion holds for fixed values of $n$, hence cannot describe
$x$ in the vicinity of $1$. 
The above formulas lead to the following large-$b$ form of $f^{\rm evol}_{\rm NS}$ in the $x$ space:
\begin{eqnarray}
&& f^{\rm evol}_{\rm NS}(x,b,Q)=x \left ( \frac{b \Lambda_{\rm QCD}}{2} \right )^{-8C_F r_0(Q_0^2,Q^2)}
 \left ( \frac{Q}{Q_0}\right )^{-\frac{4C_F}{\beta_0^{(3)}}} \nonumber \\ &&\times
\exp \left ( \left [ 2 \gamma-\frac{3}{2} \right ] r_0(Q_0^2,Q^2) + \frac{1}{b^3} r_{-3/2}(Q_0^2,Q^2) \right ) 
\nonumber \\
&&  \hspace{4cm}  {\rm for} \;\; Q^2 < \mu_c^2, \label{fnsas3}
\end{eqnarray}
and 
\begin{eqnarray}
&&f^{\rm evol}_{\rm NS}(x,b,Q)=x \left ( \frac{b}{2} \right )^{-8C_F r_0(Q_0^2,Q^2)} 
\Lambda_{\rm QCD}^{-8C_F r_0(Q_0^2,\mu_c^2)} \nonumber \\ &\times& 
\Lambda_4^{-8C_F r_0(\mu_c^2,Q^2)} \left ( \frac{\mu_c}{Q_0}\right )^{-4C_F/\beta_0^{(3)}} 
 \left ( \frac{Q}{\mu_c}\right )^{-4C_F/\beta_0^{(4)}} \nonumber \\ &\times&
\exp \left ( \left [ 2 \gamma-\frac{3}{2} \right ] r_0(Q_0^2,Q^2) 
+ \frac{1}{b^3} r_{-3/2}(Q_0^2,Q^2) \right ) 
\nonumber \\
&&  \hspace{4cm}  {\rm for} \;\; Q^2 \ge \mu_c^2, \label{fnsas4}
\end{eqnarray}
We note a few facts: the non-singlet UPD of the pion is for large $Q b$ linear in $x$ for $x$ not too
close to 1, with the slope decreasing 
with $b$ as a power law (we can neglect here the small correction due to the last term 
in Eq.~(\ref{fnsas3},\ref{fnsas4}). The exponent of $b$ is $-8C_F r_0(Q_0^2,Q^2)$. The overall constant 
is also determined. Note that linearity with $x$ at not-too-large $x$  
is seen in Fig.~\ref{Q2} for the valence quarks (dashed lines) 
at $b=10$~GeV$^{-1}$. Numerically, at $Q=2~{\rm GeV}$ we find for low $x$ that 
\mbox{$f_{\rm NS}(x, b=5~{\rm GeV}^{-1}, Q=2~{\rm GeV})=0.73 x$} and 
\mbox{$f_{\rm NS}(x, b=10~{\rm GeV}^{-1}, Q=2~{\rm GeV})=0.35 x$}, in accordance with the exact calculation of Fig.~\ref{Q2}.

The asymptotic form (\ref{fnsas3},\ref{fnsas4}) works very efficiently in practice. For the case of 
the non-singlet quarks this can be seen 
from Fig.~\ref{fig:b}, where for $Q=2$~GeV and $b > 5~{\rm GeV}^{-1}$ there is virtually no difference
between the exact numerical calculation and the asymptotic formula. 

The power-law behavior in $b$ shows that the evolution generates a rather weak behavior at large $b$. 
Numerically, at $Q=2$, $4$, and $100$~GeV, the power of $b$ is, respectively, -1.12, -1.29, and -1.75. 
This means that the large-$b$ behavior for the non-singlet quarks 
is controlled by the initial profile $F^{\rm NP}(b)$, which in chiral
quark models of the pion has an exponential fall-off, rather than by the QCD evolution. 

Now we pass to the discussion of the singlet case of Eq.~(\ref{eqS}). In the large-$Qb$ limit the leading 
part of the matrix $\Gamma_n$ of Eq.~(\ref{eqSformal}) becomes 
\begin{eqnarray}
\Gamma_n(Qb) \to \left ( \begin{array}{cc} 4C_F \log \frac{Q^2b^2}{4} & 0 \\ 
0 & 4N_c \log \frac{Q^2b^2}{4} \end{array}\right ) .
\end{eqnarray}
From this form, using methods as for the non-singlet case above, we infer that the dependence on $b$ is 
asymptotically 
\begin{eqnarray}
f_{\rm S}(x,b,Q) &\sim& b^{-8C_F r_0(Q_0^2,Q^2)}, \nonumber \\
f_{\rm G}(x,b,Q) &\sim& b^{-8N_c r_0(Q_0^2,Q^2)}. \label{powG}
\end{eqnarray}
Thus the singlet quarks fall off at the same rate as the non-singlet quarks of Eq.~(\ref{fnsas3},\ref{fnsas4}), while
the gluons drop significantly faster, as $C_f=4/3$ and $N_c=3$. 
This behavior is clearly seen in Fig.~\ref{fig:b}.
We note that due to the complications of Eq.~(\ref{eqSformal}) a more detailed analysis
yielding pre-factors, such as the one 
for the non-singlet case presented above, is more difficult in the present case,
hence we do not pursue it further here. 

We end this section with a couple of remarks concerning the observed long-range nature of the tails in $b$. 
Our initial non-perturbative profiles $F^{\rm NP}(b)$ drop exponentially, therefore suppress the tails 
generated by the evolution. This means that the large-$b$, or low-$k_\perp$ behavior is controlled 
by non-perturbative effects. 
The larger negative power in $f_G$, Eq.~(\ref{powG}), explains the the faster shrinkage of gluon
distributions in the $b$-space, or faster spreading in the $k_\perp$-space.

\section{Behavior of $f_{\rm NS}$ at $x \to 1$\label{sec:x1}}

According to standard properties of the Mellin
transform, the limit $x \to 1 $ is obtained form the large-$n$
behaviour of the anomalous dimensions. Using the asymptotic form
$B(n,m) \to \frac{\Gamma(m)}{n^m}$
in Eq.~(\ref{Beta}),
or the explicit form of the anomalous dimensions (\ref{anombNS},\ref{anomb}),
one obtains that 
\begin{eqnarray}
\gamma_{n,NS} (Q b) - \gamma_{n,NS}^{(0)} = 2 C_F \frac{b^2 Q^2}{n^2} +{\cal O}(1/n^3).
\label{gnsas}
\end{eqnarray}
We also need the large-$n$ expansion of $\gamma_{n,NS}^{(0)}$, which is
\begin{eqnarray}
\gamma_{n,NS}^{(0)}=8C_F  \left ( \log n + \gamma-3/4 +R(n) \right), \label{gnsas2}
\end{eqnarray} 
where $R(n)=\sum_{k=1}^\infty c_k n^{-k}$. Thus, according to Eq.~(\ref{nschan}), we have
the asymptotic form
\begin{eqnarray}
&&\!\!\!\!\! \frac{f^{\rm evol}_{\rm NS}(n,b,Q^2)}{f^{\rm evol}_{\rm NS}(n,b,Q_0^2)} = n^{- 8C_F r_0(Q^2,Q_0^2)} \label{fbf03} \\
&\times&  e^{- 8C_F r_0(  \gamma-3/4 +R(n)  )}  e^{2C_F b^2 r_1/n^2+{\cal O}(1/n^3)}.
\nonumber
\end{eqnarray} 
Using the initial moments (\ref{initmom}) and expanding the exponentials in Eq. (\ref{fbf03}) we 
obtain
\begin{eqnarray}
f^{\rm evol}_{\rm NS}(n,b,Q^2)&=& \frac{1}{n+1}n^{- 8C_F r_0} \left ( 1+\sum_{k=1}^\infty c'_k n^{-k} \right ) \nonumber \\ 
&\times&  \left ( 1+ 2C_F b^2 r_1/n^2+{\cal O}(1/n^3) \right ). \nonumber \\ \label{fns4}
\end{eqnarray} 
Next, we use the formula for the inverse Mellin transform 
\begin{eqnarray}
\int_C dn x^{-n} \frac{n^{-A}}{n+w} &=& (-w)^{-A} x^w \left ( 1 - \frac{\Gamma(A,-w \log \frac{1}{x}}{\Gamma(A)}\right )
\nonumber \\
& \to & \frac{x^w (\log \frac{1}{x})^{A}}{A \Gamma(A)}, \label{melex}
\end{eqnarray}
which after a straightforward algebra leads to the equation
\begin{eqnarray}
\frac{f^{\rm evol}_{\rm NS}(x,b,Q^2)}{f^{\rm evol}_{\rm NS}(x,0,Q^2)} &=& 1 - \frac{2C_F b^2 r_1 (1-x)^2}
{(1+8C_F r_0)(2+8C_F r_0)} \nonumber \\ &+& {\cal O}((1-x)^3). \label{x1}
\end{eqnarray} 
The terms with coefficients $c'_k$ do not enter at the leading order in $(1-x)$.
The behavior of Eq.~(\ref{x1}) agrees with the behaviour of Fig.~\ref{Q2},
where close to $x=1$ the departure of the curves with finite $b$ from the curve with $b=0$
becomes very slow, as it is suppressed by $(1-x)^2$.

\begin{figure}[tb]
\begin{center}
\includegraphics[width=8.6cm]{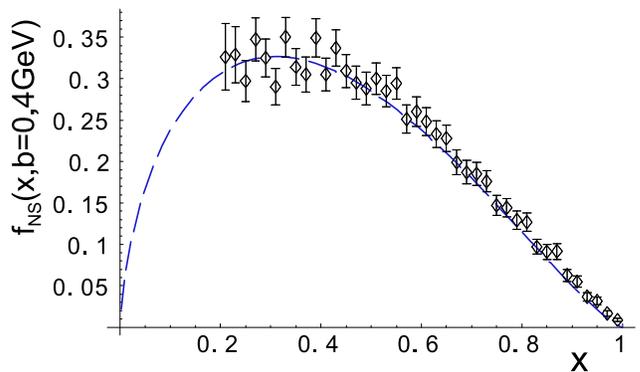}
\end{center} 
\caption{Model prediction for the {\em integrated} valence quark distribution
of the pion, evolved to from the initial condition (\ref{initmom}) to the scale of $Q=4~{\rm GeV}$, 
confronted with the E615 Drell-Yan data \cite{e615}.
The behavior at $x \to 1$ is $(1-x)^{1.29}$.}
\label{fig:e615}
\end{figure}

We also obtain that at $x \to 1$ the integrated non-singlet distribution behaves as
\begin{eqnarray}
f_{\rm NS}(x,0,Q^2) \to \frac{e^{2 C_F (3-4\gamma)r_0}}{\Gamma(1+8C_F r_0)} (1-x)^{8 C_F r_0}. \label{fNSx1} 
\end{eqnarray}
This agrees with the fact that
a function which
originally behaves at $x \to 1$ as \mbox{$f_{\rm NS}(x,0,Q_0) \to (1-x)^p$} evolves
into \cite{Peterman}
\begin{eqnarray} 
f_{\rm NS}(x,0,Q^2) \to (1-x)^{p - \frac{4 C_F }{\beta_0} \log \frac{\alpha(Q)}{ \alpha(Q_0) }}. 
\label{endpoint}
\end{eqnarray} 
In our approach the integrated function at $Q=Q_0$ has $p=0$. Numerically, we find that
\begin{eqnarray}
f_{\rm NS}(x,0,(2~{\rm GeV})^2)&\to&1.15 (1-x)^{1.13}, \nonumber \\
f_{\rm NS}(x,0,(4~{\rm GeV})^2)&\to&1.08 (1-x)^{1.29}. \label{endnum}
\end{eqnarray}
Note that although with the DGLAP evolution 
the Brodsky-Lepage counting rules for the behavior at $x \to 1$ are clearly disobeyed,
our numerical predictions agree within experimental uncertainties with 
the experimental data, including the region very close to $x=1$. 
Fig.~\ref{fig:e615} confronts our results evolved to the scale of $Q=4~{\rm GeV}$ to the
E615 experimental Drell-Yan data \cite{e615} which cover the large-$x$ region. 
In view of the simplicity of the present model, the quality of this comparison is impressive.

\section{Behavior at $x \to 0$\label{sec:x0}}

The low-$x$ behaviour of the inverse
Mellin transform is encoded in the closest singularities to the
integration line. We start with the non-singlet case of Eq.~(\ref{eqNS}).
Using the pole-residue expansion of Appendix \ref{sec:polres} we find that the 
the closest singularity is at $n=-1$, with the residue $-4C_F J_0(Q b)$.
With the help of the expression for the inverse Mellin transform,
\begin{eqnarray}
\int_{-i \infty}^{i \infty} \frac{dn}{2 \pi i} \frac{x^{-n} }{n+1} e^{\frac{a}{n+1}} = 
\left \{ \begin{array}{ll}
      x I_0(2 \sqrt{a \log \frac{1}{x}}),\;\ a \ge 0 \\
      x J_0(2 \sqrt{a \log \frac{1}{x}}),\;\ a<0     \end{array} \right . , \nonumber \\
\label{mel2}
\end{eqnarray}
we find that 
\begin{eqnarray}
f^{\rm evol}_{\rm NS}(x,b,Q^2) \to \left \{ \begin{array}{ll} 
  x I_0(2 \sqrt{C_F A \log \frac{1}{x}}),\;\ A \ge 0 \\
  x J_0(2 \sqrt{C_F A \log \frac{1}{x}}),\;\ A<0     \end{array} \right .  \nonumber \\ \label{fnsx0}
\end{eqnarray}
where 
\begin{eqnarray}
A=\int_{Q_0^2}^{Q^2}\frac{dQ^2}{2 \pi Q^2}\alpha(Q^2) J_0(Q b).
\label{Adef}
\end{eqnarray}

For the singlet case we retain the closest singularity at $n=0$ and rewrite Eq.~(\ref{eqSformal})
in the form
\begin{eqnarray}
&&\left ( \begin{array}{c} f_S(n,b,Q) \\f_G(n,b,Q) \end{array} \right ) = \label{eqSformal2} \\
&& \exp \left [ {\int_{Q_0^2}^{Q^2} \frac{dQ'^2 \alpha(Q'^2)J_0(Q'b)}{\pi Q'^2} 
\left ( \begin{array}{cc} 0 & 0 \\ 
\frac{C_F}{n} & \frac{N_c}{n} \end{array} \right )} \right] \nonumber \\ && \times 
\left ( \begin{array}{c} f_S(n,b,Q_0) \\ f_G(n,b,Q_0) \end{array} \right ), \nonumber
\end{eqnarray}
where we have used Eq.~(\ref{prS}), and could drop the ${\cal P}$ symbol since 
in the present approximation the matrix in the 
exponent does not depend on $Q'$. After some straightforward
algebra we obtain
\begin{eqnarray}
f_S(n,b,Q) &=& f_S(n,b,Q_0) , \nonumber \\
f_G(n,b,Q) &=&  f_S(n,b,Q_0) \frac{C_F}{N_c} \left ( e^{\frac{2N_cA}{n}} -1 \right ) \nonumber \\
&+& f_G(n,b,Q_0)  e^{\frac{2N_cA}{n}}.  
\end{eqnarray}
The equation for $f_S$ shows the inadequacy of retaining for this case the singularity at $n=0$ only, as
in the considered limit of $x \to 0$ the singlet quarks are controlled by the singularity at $n=-1$.  
For the case of gluons we use the initial condition (\ref{initmom}). We need the 
inverse Mellin transform
\begin{eqnarray}
&&\int_C dn x^{-n} \frac{1}{n+1} \exp \left ( {\frac{a}{n}} \right ) \nonumber \\
&=& \sum_{k=0}^\infty (-1)^k 
\left ( \frac{\log \frac{1}{x}}{a}\right )^{k/2} I_k(2 \sqrt{a \log \frac{1}{x}}) \nonumber \\
&\sim& \frac{\exp \left ( 2 \sqrt{a \log \frac{1}{x}} \right ) }{\sqrt{4 \pi \sqrt{a \log \frac{1}{x}}}
\left ( 1+\sqrt{\frac{\log \frac{1}{x}}{a}}\right ) }, \;\;\;\;\; a >0,
\label{invmeln1}
\end{eqnarray}
where in the last line we have used the asymptotic form of the Bessel functions. With this result
we find that the unintegrated 
gluon distribution at $x \to 0$ behaves as
\begin{eqnarray}
f_G(n,b,Q) \sim  \exp \left ( 2 \sqrt{2N_c A \log \frac{1}{x}} \right ), \;\; A>0, \label{DLLAg}
\end{eqnarray}
with $A$ provided in Eq.~(\ref{Adef}). If $a<0$ in Eq.~(\ref{invmeln1}), 
then the $I_k$ functions above are replaced with the $J_k$ 
functions, and the asymptotics changes the character from exponential to oscillatory:
\begin{eqnarray}
&& f_G(n,b,Q) \sim \left [ \left ( 1+\sqrt{\frac{\log \frac{1}{x}}{-A}} \right )
\cos \left (2 \sqrt{-A \log \frac{1}{x}} \right ) 
\nonumber \right . \\ && + \left .     
\left ( 1-\sqrt{\frac{\log \frac{1}{x}}{-A}} \right )
\sin \left ( 2 \sqrt{-A \log \frac{1}{x}} \right ) \right ]  \nonumber \\
&& \times \frac{1}{
\sqrt{\pi \sqrt{-A \log \frac{1}{x}}}\left ( 1-\frac{\log \frac{1}{x}}{A} \right )} , \;\;\;\; A <0 .
\label{DLLAgosc}
\end{eqnarray}

For $b=0$ the result (\ref{DLLAg}) is consistent with the double-leading-logarithmic  approximation (DLLA) 
for the DGLAP equation \cite{DLLA}, where one obtains, with constant $\alpha$,
\begin{eqnarray}
x g(x,Q)\sim \exp \left ( 2 \sqrt{ \frac{N_c}{\pi} 
\alpha \log \frac{Q^2}{Q_0^2} \log \frac{1}{x}} \right ) . \label{eq:gol}
\end{eqnarray} 
See, {\em e.g.}, the review in 
Ref.~\cite{golec}. Our formulas (\ref{fnsx0}) and (\ref{DLLAg}) are
generalizations of this behavior for the unintegrated distributions evolved with Eq.~(\ref{Keq}).

The pole-residue expansion of Appendix \ref{sec:polres} is good for
not-too-large $Q b$.  This limitation, at any fixed $x$, carries over
to Eq.~(\ref{fnsx0}) Numerically, we find that at $x=0.1$
Eq.~(\ref{fnsx0}) is valid for $Q b \leq 5$. For higher values
corrections from further residues should be included.

\section{Evolution of $\langle k_\perp^2 \rangle$\label{sec:kt}}

The average transverse momentum squared is a convenient measure of the
width of the UPD's.  Due to the factorization of the initial profile, Eq.~(\ref{factor}), $\langle
k_\perp^2 \rangle$ decomposes into two terms: the contribution from
the initial profile $F^{\rm NP}$, gives the width at the initial scale $Q=Q_0$,
and the piece $\langle k_T^2 \rangle_{\rm evol}$, entirely to the
evolution and independent of the profile $F^{\rm NP}$,
\begin{eqnarray}
\langle k_\perp^2 \rangle&=&\langle k_\perp^2 \rangle_{\rm NP}+\langle k_\perp^2 \rangle_{\rm evol}, \label{kperp} \\
\langle k_\perp^2 \rangle_{\rm NP}&=&-4 \left . \frac{d F^{\rm NP}(b)/db^2}{F^{\rm NP}(b)} \right |_{b=0}, \nonumber \\
\langle k_\perp^2 \rangle_{\rm evol}&=&-4 \left . \frac{d f^{\rm evol}(b,x,Q)/db^2}{f^{\rm evol}(b,x,Q)} \right |_{b=0}. 
\nonumber
\end{eqnarray}
The contribution $\langle k_T^2 \rangle_{\rm NP}$ has already been discussed in Sec. \ref{sec:init}, hence
here we analyze the term generated by the evolution.

Let us denote 
\begin{equation}
f_j^{(1)}(x,Q) \equiv \left . \frac{df^{\rm evol}_j(b,x,Q)}{db^2} \right | _{b=0}. \label{f1Q}
\end{equation}
In the Mellin space, the equations obtained by expanding Eq. (\ref{gamdef0}) up to first order in $b^2$
around $b=0$ read 
\begin{eqnarray}
\frac{df^{(1)}_{\rm NS}(n,Q )}{dQ^2 } & =&  -\frac{\alpha_S(Q^2)}{8 \pi Q^2}
\left ( \gamma_{n,NS}^{(0)} f^{(1)}_{\rm NS}(n,Q) \right . \nonumber \\
&+& \left . Q^2 \gamma_{n,NS}^{(1)} f_{\rm NS}(n,0,Q) \right ) , \label{df1}
\end{eqnarray}
and
\begin{eqnarray}
\frac{ d f^{(1)}_{S}(n,Q )}{dQ^2 } &=& -\frac{\alpha_S(Q^2)}{8 \pi Q^2}
\left ( \gamma_{n,qq}^{(0)} f_{S}^{(1)}(n,Q)
\right . \nonumber \\ &+& \left . \gamma_{n,qG}^{(0)} f_{G}^{(1)}(n,Q) + \right . \\ 
&& \!\!\!\!\!\! \left .  Q^2 [ \gamma_{n,qq}^{(1)} f_{S}(n,0,Q)+ \gamma_{n,qG}^{(1)} f_{G}(n,0,Q) ]  \right ), \nonumber \\
\frac{ d f^{(1)}_{G}(n,Q )}{dQ^2 } &=& -\frac{\alpha_S(Q^2)}{8 \pi Q^2}
\left ( \gamma_{n,Gq}^{(0)} f_{S}^{(1)}(n,Q) \right . \\ &+& \left .
 \gamma_{n,GG}^{(0)} f_{G}^{(1)}(n,Q)  \right . + \nonumber \\ 
&& \!\!\!\!\!\!\! \left .  Q^2  [ \gamma_{n,Gq}^{(1)} f_{S}(n,0,Q)+ \gamma_{n,GG}^{(1)} f_{G}(n,0,Q) ]  \right ) \nonumber, 
\label{eqSprim}
\end{eqnarray} 
which form a set of ordinary inhomogeneous differential equations. 
Since at the scale $Q_0$ all the width is by construction generated by the initial profile $F$, the 
initial conditions for Eq. (\ref{df1},\ref{eqSprim}) are 
\begin{equation}
f^{(1)}_{j}(n,Q_0)=0. \label{f1cond}
\end{equation}
For the non-singlet case we have the solution
\begin{eqnarray}
f^{(1)}_{\rm NS}(n,Q )&=& -
\gamma_{n,NS}^{(1)} f_{\rm NS}(n,0,Q) r_1(Q_0^2,Q^2).\nonumber \\
\label{momNS}
\end{eqnarray}
In the singlet channel we carry the analysis numerically.

\begin{figure}[tb]
\begin{center}
\epsfig{figure=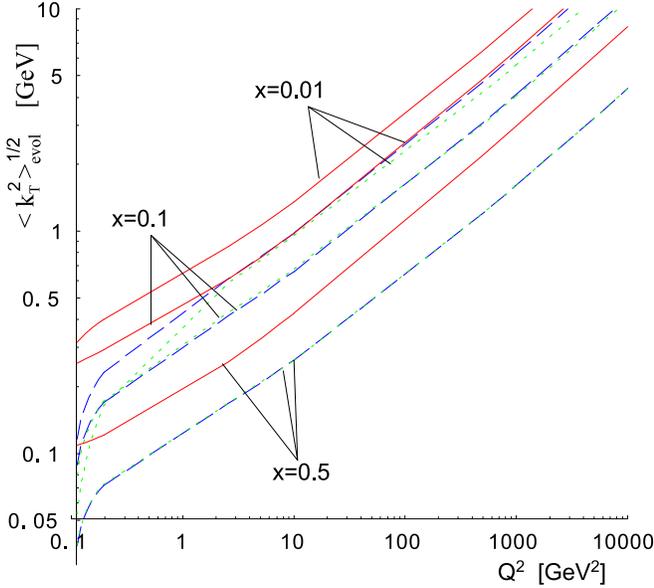,width=8.5cm}
\end{center}
\caption{ The rms transverse momenta of UPD's of the pion for
$x=0.01$, $0.1$, and $0.5$, plotted as functions of the
renormalization scale $Q^2$. Solid lines -- gluons, dashed lines --
non-singlet quarks, dotted lines -- singlet quarks.}
\label{fig:ktx}
\end{figure}

Next, we pass to the $x$-space via the numerical inverse Mellin transform. The results for 
the dynamically-generated root mean squared radius of the pion
are shown in Fig.~\ref{fig:ktx} for various values of $x$. 
In confirmation of the results of Ref.~\cite{JK3}, we note that the $k_\perp$ width 
increases with $Q$ for all parton species. The width for the gluons (solid lines) is larger than the width of 
the non-singlet (valence) quarks (dashed lines), and the singlet quarks (dotted lines).
With the log-log scales of Fig.~\ref{fig:ktx} the slopes of the plotted lines become to a good approximation equal to
one another at large $Q^2$.

With the help of previously-derived expressions for the behaviour of $f_{\rm NS}$ 
near $x=0$ and $x=1$ we may obtain the following expressions $\langle k_\perp^2 \rangle_{\rm NS}^{\rm evol}$ 
near the end-points. From Eq.~(\ref{endpoint}) we have at $x \to 0$
\begin{eqnarray}
\langle k_\perp^2 \rangle^{\rm evol}_{\rm NS} &\to& \frac{I_1(\sqrt{-4C_F r_0 \log x})}{I_0(\sqrt{-4C_F r_0 \log x})}
\sqrt{-\frac{C_F \log x}{r_0}}r_1 \nonumber \\
&\sim& \sqrt{-\frac{C_F \log x}{r_0}}r_1. \label{kperp1}  
\end{eqnarray}
At large $Q^2$ the leading behavior is
\begin{eqnarray}
\langle k_\perp^2 \rangle^{\rm evol}_{\rm NS} &\to& \sqrt{\frac{2 \beta_0^{(4)} 
C_F \log \frac{1}{x}}{\log \frac{\alpha(\mu_c^2)}{\alpha(Q^2)}}}
\frac{\alpha(Q^2)}{8 \pi} Q^2, \label{kperpas}
\end{eqnarray} 
{\em i.e.} up to the $\log \log Q^2$ corrections the spreading proceeds with $\alpha(Q^2) Q^2$.
At $x \to 1$ we find from Eq.~(\ref{x1}) that 
\begin{eqnarray}
\langle k_\perp^2 \rangle^{\rm evol}_{\rm NS} &\to& 
\frac{2C_F (1-x)^2 r_1}{(1+8C_F r_0)(2+8C_F r_0)} \label{kx1}
\end{eqnarray}
At large $Q^2$ the leading behavior is
\begin{eqnarray}
\langle k_\perp^2 \rangle^{\rm evol}_{\rm NS} &\to& 
\frac{\beta_0^{(4)2} (1-x)^2 \alpha(Q^2)}{64 \pi C_F \left [ \log \frac{\alpha(\mu_c^2)}{\alpha{Q^2}}
\right ]^2}Q^2.
\label{kx1as}
\end{eqnarray}
Again, the growth is, up to the $\log \log Q^2$ corrections, proportional to $\alpha(Q^2) Q^2$.

For the gluons the same asymptotic behavior of 
$\langle k_\perp^2 \rangle^{\rm evol}_{\rm G}$  follows from Eq.~(\ref{invmeln1}). Thus, to summarize, all 
UPD's grow at large $Q$ as $Q^2 \alpha(Q^2)$, in accordance to the behavior in Fig.~\ref{fig:ktx}.  

Interestingly, it can be noticed from Fig.~\ref{fig:ktx} that at $Q \to Q_0^+$ the $k_\perp$ width for the gluons
does not vanish. In this limit both $f^{\rm evol}_G(x,0,Q)$ and $f^{(1)}_G(x,Q)$
vanish, as is obvious from Eq.~(\ref{noglue2},\ref{f1cond}).
Thus one has a 0/0 limit. From Eq.~(\ref{Keq},\ref{df1})
with the initial condition (\ref{thetas},\ref{eqSprim}) one can easily obtain that
\begin{eqnarray}
&& \lim_{Q \to Q_0}\langle  k_\perp^2 \rangle_G^{\rm evol}=Q_0 \frac
{\int_x^1 dz\, P_{Gq}(z)\frac{(1-z)^2}{z}}{\int_x^1 dz\, P_{Gq}(z)\frac{1}{z}} \label{limG}\\
&&=Q_0 \frac{x^4-6x^3+21x^2-18 x \log x-10x-6}{3(x^2-2x \log x +x -2)}, \nonumber
\end{eqnarray}  
which is positive for $x \in [0,1)$ and equal 0 for $x=1$.
On the other hand, since for the quarks $f^{\rm evol}_{NS,S}(x,0,Q) \neq 0$ 
$\langle k_\perp^2 \rangle^{\rm evol}_{NS,S}$ vanish at $Q_0$.

In phenomenological applications it is sometimes useful to have a
simple formula characterizing the discussed behavior.  In the range
$2~{\rm GeV}^2 < Q^2 < 10000~{\rm GeV}^2$ and $0.005 < x < 0.8$ the
following simple-minded interpolating formula works to within a few per cent:
\begin{eqnarray}
\left ( \langle k_\perp^2 \rangle_i^{\rm evol} \right )^{1/2} \!\!\!=
A_i(\log \frac{1}{x}) \left( \frac{Q^2}{\Lambda_{\rm QCD}^2} \right )^
{0.35+0.004 \log \frac{Q^2}{\Lambda_{\rm QCD}^2} } \!\!\!\!\! ,
\nonumber \\ \label{approx}
\end{eqnarray}
where $i=$NS, S, or G, and
\begin{eqnarray}
A_{\rm NS}(y)&=&-0.017 y^{1/2}+0.113 y -0.057 y^{3/2}+ 0.010 y^2, \nonumber \\
A_{S} (y)&=&-0.021 y^{1/2}+0.120 y -0.059 y^{3/2}+ 0.009 y^2, \nonumber \\
A_{G} (y)&=&-0.016 y^{1/2}+0.150 y -0.075 y^{3/2}+ 0.011 y^2. \nonumber \\
\label{Apar}
\end{eqnarray}
The power of $Q^2$ of 0.35 in Eq.~(\ref{approx}), rather than
$1/2$ corresponding to  $\langle k_\perp^2 \rangle_i^{\rm evol} \sim Q^2 \alpha(Q^2)$, 
compensates, in the chosen range for $Q$, for the logarithmic corrections.
We note that Eq.~(\ref{approx}) holds for the pion with the initial conditions (\ref{initmom}) provided
by the chiral quark models.  

\section{Formal limits for other initial conditions\label{sec:general}}

Certain formal results listed in this paper, such as the formulas for the 
non-singlet quarks, (\ref{fnsas3}, \ref{fnsas4}, \ref{fNSx1}, \ref{endpoint}, \ref{fnsx0}, \ref{kperp1},%
\ref{kperpas}), are specific to the
evolution with the
initial condition following from the chiral quark models, Eq.~(\ref{initx}, \ref{initmom}). 
However, these results can be easily generalized. Note that most of the popular   
parameterizations of initial conditions, such as those of Refs.~\cite{GRS,GRV}, involve
linear combinations of $x^\alpha (1-x)^\beta$. It is understood that the factorization
in the initial condition between $x$ and $b$ variable holds, as assumed in Ref.~\cite{JK3}.

For the case of large-$b$ asymptotics, the relevant
formula is (\ref{lapy}). With the initial condition $x^\alpha (1-x)^\beta$ it becomes
\begin{eqnarray}
\int_C \frac{dn}{2\pi i} x^{-n} \frac{e^{y_0+y_1 n}}{n+\alpha} = x^\alpha e^{y_0-\alpha y_1} 
\Theta \left (y_1+\log \frac{1}{x} \right ),
\label{lapygen}
\end{eqnarray}
and expressions (\ref{fnsas3},\ref{fnsas4}) are modified accordingly, with $x$ replaced by 
$x^\alpha$, and $y_1$ in the exponent multiplied by $\alpha$. The formulas 
(\ref{powG}) remain valid. Therefore the power-law behavior at large $b$ is independent of the 
initial condition, and is $b^{-8C_F r_0(Q^2_0,Q^2)}$ for the quarks and $b^{-8N_c r_0(Q^2_0,Q^2)}$ 
for the gluons. 

In the limit of $x \to 1$ the only difference is the appearance of the extra power of $\beta$ in the 
in Eq.~(\ref{fNSx1}). Thus, the UPD's for finite $b$ approach the $b=0$ case of the integrated distributions 
as $(1-x)^2$.  

In the limit of $x \to 0$ we need generalizations of the Mellin transforms of Eq.~(\ref{mel2},\ref{invmeln1}) 
for $\alpha \neq 1$. These are
\begin{eqnarray}
&&\int_C dn x^{-n} \frac{1}{n+\alpha} \exp \left ( {\frac{a}{n+1}} \right ) \\
&=& x \sum_{k=0}^\infty (-1)^k (\alpha-1)^k
\left ( \frac{\log \frac{1}{x}}{a}\right )^{k/2} I_k(2 \sqrt{a \log \frac{1}{x}}) \nonumber \\
&\sim&  \frac{ x \exp \left ( 2 \sqrt{a \log \frac{1}{x}} \right ) }{(\alpha-1)\sqrt{4 \pi \sqrt{a \log \frac{1}{x}}}
\sqrt{\frac{\log \frac{1}{x}}{a}}}, \;\;\;\; \alpha \neq 1, \;a>0, \nonumber
\label{invmeln2}
\end{eqnarray}
and
\begin{eqnarray}
&&\int_C dn x^{-n} \frac{1}{n+\alpha} \exp \left ( {\frac{a}{n}} \right ) \\
&=& \sum_{k=0}^\infty (-1)^k \alpha^k
\left ( \frac{\log \frac{1}{x}}{a}\right )^{k/2} I_k(2 \sqrt{a \log \frac{1}{x}}) \nonumber \\
&\sim& \frac{\exp \left ( 2 \sqrt{a \log \frac{1}{x}} \right ) }{\alpha \sqrt{4 \pi \sqrt{a \log \frac{1}{x}}}
\sqrt{\frac{\log \frac{1}{x}}{a}}}, \;\;\;\; \alpha \neq 0,\; a>0 . \nonumber
\label{invmeln3}
\end{eqnarray}
The analogs of Eq.~(\ref{fnsx0},\ref{DLLAg},\ref{kperp1},\ref{kperpas}) follow straightforwardly. 
In particular, the $Q^2 \alpha(Q^2)$ large-$Q$ behavior for all parton distributions is preserved.

\section{Conclusions\label{sec:con}}

We have presented a new method of solving the Kwieci\'nski equations for the 
leading-order QCD evolution of  
unintegrated parton distributions. The method is based on the Mellin transform
and parallels the standard analysis of the DGLAP equations. 
Our main results are as follows:

\begin{enumerate}
\item We have found analytic forms of the $b$-dependent anomalous dimensions, expressed 
through hypergeometric functions, which allowed us 
to study formal aspects of the equations and their solutions, {\em e.g.} the 
asymptotic forms of the evolution-generated UPD's at large $b$, or at $x\to 0$ and $x \to 1$.
We have also demonstrated that the proposed numerical method is fast and stable. 

\item The numerical work can be simplified if low-$b$ or large-$b$ expansions are used.

\item At large $b$ the evolution-generated $b$-dependent UPD's exhibit power-law fall-off, with the 
magnitude of the exponents 
growing with the probing scale 
$Q$, {\em cf.} Eq.~(\ref{fnsas3},\ref{powG}). The fall-off is steeper for the gluons than for the quarks.

\item At $x \to 0$ we have found generalizations of the DLLA behavior, {\em cf.} Eq.~(\ref{fnsx0},\ref{DLLAg}.) 
We have also shown that for large $b$ the solution for the valence UPD of the pion
grows linearly with $x$ for not too large $x$, and the slope decreases with $b$ as a power law. 

\item At $x \to 1$ the evolution-generated $b$-dependent UPD's approach the integrated distributions as $(1-x)^2$.

\item Our numerical results fully confirm the finding of Ref.~\cite{JK3},
where a different numerical method was used. We find the spreading of
the $k_\perp$ distributions with the probing scale $Q$, with the
effect strongest for gluons and increasing with decreasing $x$. We
have also shown that the widths $\langle k_\perp^2 \rangle_i^{\rm evol}$ in all
channels $i$ increase at large $Q^2$ as $Q^2 \alpha(Q^2)$. 

\item For practical purposes in possible phenomenological applications, we have
parameterized $\langle k_\perp^2 \rangle_i^{evol}$ with a simple formula
which works with accuracy of a few percent.

\end{enumerate}

Although the specific study of this paper was devoted to the pion with
the initial condition following from the chiral models, and several of
the more detailed analytic formulas were specific to this case, the developed method
is general and can be applied to any initial form of the data. In
particular, it can be used with the GRS \cite{GRS} or GRV \cite{GRV}
parameterization supplied by a profile in $b$, such as already studied
in Ref.~\cite{JK3}. The formal results of Sec.~\ref{sec:general} are general for 
a wide class of initial conditions, suitable for both the pion and the nucleon.

\begin{acknowledgments}
We thank Krzysztof Golec-Biernat and Antoni Szczurek for useful
discussions.  One of us (WB) is grateful to Andrzej Horzela for
pointing out the impressive collection of practical mathematical
knowledge at {\tt http://www.mathworld.com}~\cite{mathworld}.  Support
from DGI and FEDER funds, under contract BFM2002-03218 and by the
Junta de Andaluc\'\i a grant no. FM-225 and EURIDICE grant number
HPRN-CT-2003-00311 is acknowledged.  Partial support from the Spanish
Ministerio de Asuntos Exteriores and the Polish State Committee for
Scientific Research, grant number 07/2001-2002 is also gratefully
acknowledged.
\end{acknowledgments}

\appendix
\section{Elements of the perturbative QCD evolution\label{pQCD}}
We use the LO QCD evolution with three active flavors up to the scale $\mu_c^2=4~{\rm GeV}^2$ 
and four active flavors above. Therefore $\alpha=g^2 / (4\pi)$ is given by
\begin{eqnarray} 
\alpha (Q^2) &=& \frac{4\pi}{\beta_0^{(3)}\log \left(\frac{Q^2}{\Lambda_{QCD}^2}\right) }, \;\;\;\;\;\; Q^2 \leq \mu_c^2,
\nonumber \\
\alpha (Q^2) &=& \frac{4\pi}{\beta_0^{(4)}\log \left(\frac{Q^2}{\Lambda_{4}^2}\right) }, \;\;\;\;\;\;\;\;\;\; Q^2 > \mu_c^2,
\nonumber \\
\Lambda_4 &=& \mu_c 
\left ( \frac{\Lambda_{QCD}}{\mu_c} \right )^{\frac{\beta_0^{(3)}}{\beta_0^{(4)}}}, 
\label{eq:alpha}
\end{eqnarray}
with $\beta_0^{(N_f)}=11-2 N_f/3$ for $N_c=3$, 
where $N_f$ and $N_c$ denote the number of flavors and colors, respectively. 
Along this paper we take 
\begin{eqnarray}
\Lambda_{\rm QCD}=226~{\rm MeV},
\end{eqnarray}
as was done
in Refs.~\cite{DavArr1,DavArr2,Weigel,SQM}.
The value of the scale $\Lambda_4$ ensures matching at $Q^2=\mu_c^2$. Numerically, $\beta_0^{(3)}=9$, 
$\beta_0^{(4)}=25/3$, and $\Lambda_4=189~{\rm MeV}$.

The functions $r_k$, defined in Eq.~(\ref{pQCD}), have the explicit form 
\begin{eqnarray}
r_0(Q_0^2,Q^2)  &=&
\frac{1}{2 \beta_0^{(3)}} 
\log \left ( \frac{\log(Q^2/\Lambda_{\rm QCD}^2)}{\log(Q_0^2/\Lambda_{\rm QCD}^2)} \right ) 
\nonumber \\ &=& \frac{1}{2 \beta_0^{(3)}} 
\log \frac{\alpha(Q_0^2)}{\alpha(Q^2)}
 ,\;\;\;\;\;\;Q^2 \leq \mu_c^2,
\nonumber
\end{eqnarray}
\begin{eqnarray}
r_0(Q_0^2,Q^2)  &=& r_0(Q_0^2,\mu_c^2)+  
\frac{1}{2 \beta_0^{(4)}} 
\log \left ( \frac{\log(Q^2/\Lambda_4^2)}{\log(\mu_c^2/\Lambda_4^2)}\right )
\nonumber \\ &=& r_0(Q_0^2,\mu_c^2) + \frac{1}{2 \beta_0^{(4)}} 
\log   \frac{\alpha(\mu_c^2)}{\alpha(Q^2)}, \label{r0ap}\\
&& \hspace{42mm} Q^2 > \mu_c^2, \nonumber
\end{eqnarray}
and for $k \neq 0$
\begin{eqnarray}
r_k(Q_0^2,Q^2)  &=&
\frac{\Lambda_{\rm QCD}^{2k}}{2 \beta_0^{(3)}} 
\left [ {\rm Li}\left ( \frac{Q^{2k}}{\Lambda_{\rm QCD}^{2k}} \right ) - 
{\rm Li}\left ( \frac{Q_0^2}{\Lambda_{\rm QCD}^2} \right ) \right ], 
\nonumber \\ && \hspace{30mm} Q^2 \leq \mu_c^2, \nonumber 
\end{eqnarray}
\begin{eqnarray}
r_k(Q_0^2,Q^2)  &=& r_k(Q_0^2,\mu_c^2)   +  
\frac{\Lambda_4^{2k}}{2 \beta_0^{(4)}}  \label{rkap} \\ &\times&  
\left [ {\rm Li}\left ( \frac{Q^{2k}}{\Lambda_4^{2k}} \right ) - 
{\rm Li}\left ( \frac{\mu_c^2}{\Lambda_4^2} \right ) \right ], \;\;\;\;\;\;Q^2 > \mu_c^2. 
\nonumber
\end{eqnarray}
Above we have used the indefinite integral
\begin{eqnarray}
\int \frac{d Q^2 Q^{2k}}{Q^2 \log(Q^2/\Lambda^2)} &=& 
\Lambda^{2k} {\rm Li} \left( \frac{Q^{2k}}{\Lambda^{2k}} \right), \;\; k=1,2,\dots, \nonumber \\
\label{lieq}
\end{eqnarray} 
where the logarithmic integral is 
\begin{eqnarray}
{\rm Li} (x) = \int_0^x dt / \log t. \label{lidef}
\end{eqnarray}
At large $Q^2$ 
\begin{eqnarray}
\Lambda^{2k} {\rm Li} \left( \frac{Q^{2k}}{\Lambda^{2k}} \right)=
Q^{2k} \left ( \frac{1}{k \log \frac{Q^2}{\Lambda^2}} + \frac{1}{k^2 \log^2 \frac{Q^2}{\Lambda^2}}+ \dots   \right ).
\nonumber \\ \label{lamas}
\end{eqnarray} 

The functions $P_{ab}(z)$ are the LO splitting functions corresponding to real 
emission, {\em i.e.}:
\begin{eqnarray}
P_{qq}(z)&=&C_F{1+z^2\over 1-z} , \label{Psplit} \\
P_{qG}(z)&=&N_f[z^2+(1-z)^2] ,\nonumber \\
P_{Gq}(z)&=&C_F{1+(1-z)^2\over z} , \nonumber \\
P_{GG}(z)&=&2N_c\left[{z\over 1-z}+{1-z\over z} + z(1-z)\right] , \nonumber
\label{split}
\end{eqnarray}
with $C_F=(N_c^2-1)/(2N_c)=4/3$.

\section{$b$-dependent anomalous dimensions\label{anomdim}}

We introduce 
\begin{equation}
u=\frac{Q^2 b^2}{4},
\end{equation}
as well as the anomalous dimensions for the $b=0$ case of the DGLAP equations, where for the non-singlet
we have
\begin{eqnarray}
\gamma_{n,NS}^{(0)} &=& 2C_F \left( -3 + \frac{2}{1 + n} + 
\frac{2}{2 + n} + 4\,H_n \right) , \label{anom0NS}
\end{eqnarray}
while for the singlet
\begin{eqnarray}
\gamma_{n,qq}^{(0)} &=& \gamma_{n,NS}{(0)} , \label{gam0} \\
\gamma_{n,qG}^{(0)} &=& - 4 N_f \,\left( \frac{1}{1 + n} - \frac{2}{2 + n} + \frac{2}{3 + n} \right),
\nonumber \\
\gamma_{n,Gq}^{(0)} &=& -4 C_F \left( \frac{2}{n} - \frac{2}{1 + n} + 
\frac{1}{2 + n} \right), \label{anom0} \nonumber \\
\gamma_{n,GG}^{(0)} &=& 2 N_c\,\left( -3 - \frac{4}{n} + \frac{8}{1 + n} - 
\frac{4}{2 + n} + \frac{4}{3 + n} \right . \nonumber \\
&+& \left . 4\,H_n \right ) . \nonumber
\end{eqnarray}
The symbol $H_n$ denotes the harmonic number 
\begin{equation}
H_n=\sum_{k=1}^n \frac{1}{k} = \frac{\Gamma' (n+1)}{\Gamma(n+1)} + \gamma, \label{Hn}
\end{equation} 
which is a meromorphic function in the complex $n$ 
variable, with poles located at negative
integers $n= -1, -2, -3, ...$ and residues equal to $-1$.

Below we list the  anomalous dimensions for the moments of the unintegrated parton distributions in the $b$ space,
defined in Eq. (\ref{gammadef}). The formulas follow from the basic analytic integral
\begin{eqnarray}
&& \frac{\Gamma(2 + \mu + \nu)} {\Gamma(1 + \mu)\,\Gamma(1 + \nu)} 
\int_0^1 dy y^\mu (1-y)^\nu J_0\left ( 2 \sqrt{u} y  \right ) )= \nonumber \\  && 
   {}_2F_3(\frac{1+\mu}{2},\frac{2+\mu}{2};
     1,\frac{2+\mu + \nu}{2},\frac{3+\mu+\nu}{2};-u)
\nonumber \\ \label{basint}
\end{eqnarray}
and relations among the generalized hypergeometric functions. For the non-singlet case we have 
\begin{widetext}
\begin{eqnarray}
\gamma_{n,NS}(Q b)&=&\gamma_{n, NS}^{(0)}+
\frac{4 C_F}{\left( 1 + n \right) \,\left( 2 + n \right) }
\left [ -3 - 2\,n + 2\,\left( 2 + n \right) \,
{}_1F_2(\frac{1}{2};\frac{2 + n}{2},\frac{3 + n}{2};-u) \right .\nonumber \\
&-& \left . 
  {}_1F_2(\frac{3}{2}; \frac{3 + n}{2},\frac{4 + n}{2};-u) +
 2u\,{}_3F_4(1,1,\frac{3}{2}; 2,2,\frac{3 + n}{2},\frac{4 + n}{2}; -u)\right ],
\label{anombNS}
\end{eqnarray}
whereas for the singlet case
\begin{eqnarray}
\gamma_{n,qq}(Q b)&=&\gamma_{n,NS}(Q b),
\nonumber 
\end{eqnarray}
\begin{eqnarray}
\gamma_{n,qG}(Q b)&=&\gamma_{n,qG}^{(0)}+
\frac{4 N_f}{\left( 1 + n \right) \left( 2 + n \right) 
\left( 3 + n \right) \left( 4 + n \right)  \left( 5 + n \right) }
\left [ -\left( \left( 4 + n \right) \,\left( 5 + n \right) \,
     \left( -4 - n\,\left( 3 + n \right)  
\nonumber \right . \right . \right . \\ &+&
\left . \left . \left . \left( 2 + n \right) \,\left( 3 + n \right) \,  
      {}_2F_1( \frac{1}{2}; \frac{2 + n}{2},\frac{3 + n}{2};-u) - 
       2\,\left( 3 + n \right) \,{}_1F_2(\frac{3}{2}; \frac{3 + n}{2},\frac{4 + n}{2}; -u)
\nonumber \right . \right . \right . \\ &+& \left . \left . \left .
       4\,{}_1F_2(\frac{3}{2}; \frac{4 + n}{2},\frac{5 + n}{2};-u) \right)  \right)  + 
  24\,u\,{}_1F_2(\frac{5}{2}; \frac{6 + n}{2},\frac{7 + n}{2};-u)\right ],
\nonumber 
\end{eqnarray}
\begin{eqnarray}
\gamma_{n,Gq}(Q b)&=&\gamma_{n,Gq}^{(0)}+ \frac{4 C_F}{n\,\left( 1 + n \right) \,\left( 2 + n \right) 
\,\left( 3 + n \right) \,\left( 4 + n \right) }
\nonumber \\ &\times& \left [   -\left( \left( 1 + n \right) \,\left( 2 + n \right) \,\left( 3 + n \right) 
\,\left( 4 + n \right) \,
     {}_1F_2(\frac{1}{2}; \frac{1 + n}{2},\frac{2 + n}{2}; -u) \right) 
\right .  \nonumber \\ &+& \left . 
  \left( 3 + n \right) \,\left( 4 + n \right) \,\left( 4 + n\,\left( 3 + n \right)  - 
     2\, {}_1F_2(\frac{3}{2};\frac{3 + n}{2},\frac{4 + n}{2};-u) \right)  + 
  12\,u\,{}_1F_2(\frac{5}{2};\frac{5 + n}{2},\frac{6 + n}{2};-u)       \right ] ,
\nonumber 
\end{eqnarray}
\begin{eqnarray}
\gamma_{n,GG}(Q b)&=&\gamma_{n,GG}^{(0)}+ 
8 N_c \left [ \frac{1}{n} - \frac{2}{1 + n} + \frac{1}{2 + n} - \frac{1}{3 + n} + 
  \frac{{}_1F_2(\frac{1}{2}; 1 + \frac{n}{2},\frac{3}{2} + \frac{n}{2};-u)}{1 + n} - 
  \frac{{}_1F_2(\frac{3}{2};1 + \frac{n}{2},\frac{3}{2} + \frac{n}{2};-u)}{n + n^2} 
\nonumber \right . \\ &-& \left . 
  \frac{{}_1F_2(\frac{3}{2};\frac{3}{2} + \frac{n}{2},2 + \frac{n}{2};-u)}
{\left( 1 + n \right) \,\left( 2 + n \right) } + 
  \frac{2\,{}_1F_2(\frac{3}{2};2 + \frac{n}{2},\frac{5}{2} + \frac{n}{2};-u)}
   {\left( 1 + n \right) \,\left( 2 + n \right) \,\left( 3 + n \right) } - 
  \frac{12\,u\,{}_1F_2(\frac{5}{2}; 3 + \frac{n}{2},\frac{7}{2} + \frac{n}{2};-u)}
   {\left( 1 + n \right) \,\left( 2 + n \right) \,\left( 3 + n \right) 
\,\left( 4 + n \right) \,\left( 5 + n \right) } \right . \nonumber \\ &+& \left . 
  \frac{u\,{}_3F_4(1,1,\frac{3}{2};2,2,\frac{3}{2} + \frac{n}{2},2 + \frac{n}{2};-u)}
   {\left( 1 + n \right) \,\left( 2 + n \right) }  \right ] . \label{anomb}
\end{eqnarray}
\end{widetext}
One may verify that the analyticity properties in $n$ of the anomalous dimensions (\ref{anombNS},\ref{anomb})
are the same as for the $b=0$ case of (\ref{anom0NS},\ref{anom0}).

\section{Expansion of anomalous dimensions at low $b Q$\label{app:lowb}}

We may expand in the anomalous dimensions in powers of $Q^2 b^2$, 
\begin{eqnarray}
\gamma_{n,j}(Q b) =\gamma_{n,j}^{(0)}+\gamma_{n,j}^{(1)} Q^2 b^2 + \dots, \label{gambexp}
\end{eqnarray}
which yields
\begin{eqnarray}
\gamma_{n,NS}^{(1)} &=&  \gamma_{n,qq}^{(1)} = \frac{2 C_F(n^2+5n+7)}{(n+1)(n+2)(n+3)(n+4)}, \nonumber \\
\gamma_{n,qG}^{(1)} &=& \frac{2 N_f (n^2+3n+14)}{(n+1)(n+2)(n+3)(n+4)}, \nonumber \\
\gamma_{n,Gq}^{(1)} &=& \frac{2 C_F(n^2+7n+24)}{(n+1)(n+2)(n+3)(n+4)}, \nonumber \\
\gamma_{n,GG}^{(1)} &=& \frac{2N_c [n(n+5)(n^2+5n+16)+120]}{n(n+1)n+2)(n+3)(n+4)(n+5)}. \label{g1}
\end{eqnarray}
More generally, introducing the Euler Beta function, 
${\sf B}(x,y)= \Gamma(x) \Gamma(y) / \Gamma(x+y)$, and applying 
the series expansion of the Bessel function, 
\begin{equation}
J_0 (x) = \sum_{k=0}^\infty \frac{(-x^2)^k}{2^k k!^2},  \label{besexp}
\end{equation}
we arrive at the expansion formulas
\begin{eqnarray}
&&\!\!\!\!\!\! \gamma_{n,NS} ( Q b ) = \gamma_{n,qq}( Q b) =\gamma_{n,qq}^{(0)} \nonumber \\
&&-C_F\sum_{k=1}^\infty \frac{(-Q^2 b^2 )^k 4^{1-k}}{k!^2 } \left[ {\sf B} (2k,n+1) + {\sf B}
 (2k,n+3) \right] \nonumber
\end{eqnarray}
\begin{eqnarray}
&&\!\!\!\!\!\! \gamma_{n,qG} ( Q b ) = \gamma_{n,qG}^{(0)}
 -N_f \sum_{k=1}^\infty \frac{(-Q^2 b^2 )^k \, 4^{1-k}}{k!^2 } \nonumber \\
&&\times \left[ {\sf
 B} (2k+1,n+1) -2 {\sf B} (2k+1,n+2) \right . \nonumber \\
&& \;\; \left . + 2 {\sf B} (2k+1,n+3) \right] \nonumber 
\end{eqnarray}
\begin{eqnarray}
&&\!\!\!\!\!\! \gamma_{n,Gq} ( Q b ) = \gamma_{n,Gq}^{(0)} -C_F\sum_{k=1}^\infty
\frac{(-Q^2 b^2 )^k 4^{1-k}}{k!^2 } \nonumber \\
&& \times  \left[ 2 {\sf B} (2k+1,n) -2 {\sf
B} (2k+1,n+1) \right . \nonumber \\
&& \;\; \left . + {\sf B} (2k+1,n+2) \right] \label{Beta}
\end{eqnarray}
\begin{eqnarray}
&&\!\!\!\!\!\! \gamma_{n,GG} ( Q b ) = \gamma_{n,GG}^{(0)}
 -2 N_c\sum_{k=1}^\infty \frac{(-Q^2 b^2 )^k 4^{1-k}}{ k!^2 } \nonumber \\
&& \times \left[ {\sf
 B} (2k,n+2) + {\sf B} (2k+2,n) +  {\sf B} (2k+2,n+2) \right] . \nonumber 
\label{ge}
\end{eqnarray} 


\section{Asymptotics of the anomalous dimensions at large $b Q$ \label{app:largeb}}

We may use the asymptotic forms of the generalized hypergeometric functions appearing in
Eq.~(\ref{anombNS},\ref{anomb}). One has \cite{hyp12} 
\begin{eqnarray}
&& \!\!\!\!\!\!\! {}_1F_2(a_1;b_1,b_2;-\frac{Q^2 b^2}{4})=\frac{\Gamma(b_1)\Gamma(b_2)}
{\Gamma(b_1-a_1) \Gamma(b_2-a_1)} \left ( \frac{4}{Q^2 b^2} \right )^{a_1} \nonumber \\
&& + \frac{\Gamma(b_1) \Gamma(b_2)}{\sqrt{\pi}\Gamma(a_1)} 
\left \{ \cos(Q b -\pi c_1) \right . \nonumber \\ && 
\left . \hspace{2cm} +\frac{c_2}{Q b} \sin(Q b-\pi c_1) \right \}
\left ( \frac
{4}{Q^2 b^2} \right )^{c_1}+\dots , \nonumber \\
&&c_1=\frac{1}{2} \left ( b_1+b_2-a_1-\frac{1}{2} \right ), \label{1F2} \\
&&c_2=\frac{1}{8} \left ( 12 a_1^2-8(b_1+b_2+1)a_1 
-4(b_1-b_2)^2 \right . \nonumber \\ &&
\left . \hspace{2cm} +8(b_1+b_2)-3 \right ), \nonumber
\end{eqnarray}
and \cite{hyppq}
\begin{eqnarray}
&&{}_3F_4 \left ( 1,1,\frac{3}{2};2,2,\frac{n+3}{2},\frac{n+4}{2};-\frac{Q^2 b^2}{4} \right ) =
\nonumber \\
&& \frac{4(n+1)(n+2)}{n Q^3 b^3} \left ( Q b [ n \log \frac{Q b}{2} - n\psi^{0}(n) -1 ]+n^2 
\right ) \nonumber \\
&& + \frac{8 \Gamma(n+3)}{\sqrt{2\pi}} \cos \left ( Q b -
\frac{2n+3}{4}\pi \right ) (Q b)^{-\frac{n}{2} -\frac{7}{2}} +\dots  
\nonumber \\ \label{3F4}
\end{eqnarray} 
Then the following asymptotic expansions for the $b$-dependent anomalous dimensions hold:
\begin{eqnarray}
&& \gamma_{\rm NS}(n,Q b)=\gamma_{qq}(n,Q b)=4 C_F \left ( \log \frac{Q^2 b^2}{4}+ 2 \gamma - \frac{3}{2} \right .\nonumber \\
&& \left . + \frac{2n+2}{Q b} + \frac{\Gamma(n+1)(Q b)^{-n-\frac{5}{2}}}{4 \sqrt{2\pi}} \right . \nonumber \\
&& \left . \times \left [ 24 b Q \cos (\frac{2n+3}{4}\pi - Q b) \right . \right . \nonumber \\
&& \left . \left . - (12n+13) \sin (\frac{2n+3}{4}\pi - Q b) \right ] \right )+\dots \label{asqq}
\end{eqnarray}
\begin{eqnarray}
&& \gamma_{qG}(n,Q b)=4N_f \left ( - \frac{1}{Q b} + \frac{\Gamma(n+1)(Qb)^{-n-\frac{5}{2}}}{8 \sqrt{\pi}} 
\right . \nonumber \\ 
&& \left . \times \left [ (-12n+8Qb-11)\cos(\frac{n\pi}{2}-Qb) \right . \right . \nonumber \\
&& \left . \left . +
 (12n+8Qb+11)\sin(\frac{n\pi}{2}-Qb) \right ] \right ) + \dots \label{asqG}
\end{eqnarray}
\begin{eqnarray}
&& \gamma_{Gq}(n,Q b)=4C_F \left ( - \frac{1}{Q b} + \frac{\Gamma(n)(Qb)^{-n-\frac{3}{2}}}{4 \sqrt{\pi}} 
\right . \nonumber \\ 
&& \left . \times \left [ (4n+8Qb-1)\cos(\frac{n\pi}{2}-Qb) \right . \right . \nonumber \\
&& \left . \left . +
 (4n-8Qb-1)\sin(\frac{n\pi}{2}-Qb) \right ] \right ) + \dots \label{asGq}
\end{eqnarray}
\begin{eqnarray}
&& \gamma_{GG}(n,Q b)=\frac{4N_f}{3} + 4N_c 
\left ( \log \frac{Q^2 b^2}{4} + 2 \gamma - \frac{11}{6} \right . \nonumber \\
&& \left . + \frac{2n+2}{Q b} -\frac{4 \Gamma(n) (Qb)^{-n -\frac{1}{2}}}{\sqrt{2\pi}} 
\cos (\frac{2n+1}{4}\pi-Qb) \right. \nonumber \\ 
&& \left . + \frac{[\Gamma(n)-20\Gamma(n+1)] (Qb)^{-n -\frac{3}{2}}}{4\sqrt{\pi}}
\right . \nonumber \\
&& \left . \times  
[\cos (\frac{n}{2}\pi-Qb)+\sin (\frac{n}{2}\pi-Qb)]   \right )+\dots \label{asGG}
\end{eqnarray}
The ellipses denote terms subleading in $\frac{1}{Q b}$. The above
formulas assume that $n$ is kept fixed. In actual applications, such
as numerical programming of the generalized hypergeometric functions,
it is practical to switch from the general formulas
(\ref{anombNS},\ref{anomb}) to the asymptotic expressions
(\ref{asqq},\ref{asGG}) when $Q b \ge 10 |n|$.

\section{Pole-residue expansion of the anomalous dimensions\label{sec:polres}}

For $ b \neq 0 $ the analytic
structure of the $b-$dependent anomalous dimensions remains the
same as for $b=0$.
This can be seen by expanding the Bessel function in the
integrand of Eq.~(\ref{gammadef}) as a power series around $z=0$, which yields
\begin{eqnarray}
&& \gamma_{n,ab} (Q b ) = \label{pow} \\ && - 4 \int_0^1
d z  \sum_{k=0}^\infty \frac1{k!} \left [ J_0^{(k)} (Qb) (-Qb)^{k} z^{n+k} -1 \right ] P_{ab}(z). \nonumber
\end{eqnarray} 
Applying the trick
\begin{eqnarray}
1 = J_0(0) = -4
\sum_{k=0}^\infty \frac1{k!} J_0^{(k)} (Qb) (-Qb)^k, \label{sumJ}
\end{eqnarray} 
we find the expansion involving index-shifted anomalous dimensions at $b=0$, namely 
\begin{eqnarray}
\gamma_{n,ab} ( Q b ) &=& 
\sum_{k=0}^\infty  \frac{(-Qb)^k}{k!} J_0^{(k)} (Qb) 
\gamma_{n+k,ab} ( 0) . \nonumber \\ \label{gaJ}
\end{eqnarray} 
Using the explicit expression for the anomalous dimension this series
may be rewritten as a pole-residue expansion
\begin{eqnarray}
\gamma_{n,ab} ( Q b )  = 
\sum_{k=0}^\infty \frac{R_k^{ab} (Qb)}{n+k}  .
\label{eq:pole_res}
\end{eqnarray} 
In practice this means that the Mellin contour used in the case of $b=0$
can be used in the $b\neq 0$ case as well. 
In the non-singlet case the first few residues read 
\begin{eqnarray}
R_1^{\rm NS} (A) &=& -4C_F J_0(A), \label{prNS} \\
R_2^{\rm NS} (A) &=& -4C_F (J_0(A)+AJ_1(A)), \nonumber  \\
R_3^{\rm NS} (A) &=& -2C_F (-(A^2-4) J_0(A)+3A J_1(A) ), \nonumber 
\label{eq:res}
\end{eqnarray}
while in the singlet channel $R_i^{qq}=R_i^{\rm NS}$ and
\begin{eqnarray}
R_1^{qG} (A) &=& -4N_F J_0(A), \nonumber \\
R_2^{qG} (A) &=& -4N_F (-2J_0(A)+AJ_1(A)), \nonumber \\ 
\dots \nonumber \\
R_0^{Gq} (A) &=& -8C_F J_0(A), \nonumber \\
R_1^{Gq} (A) &=& -8C_F (-J_0(A)+AJ_1(A)), \nonumber \\ 
\dots \nonumber \\
R_0^{GG} (A) &=& -8N_c J_0(A), \nonumber \\
R_1^{GG} (A) &=& -8N_c (-J_0(A)+AJ_1(A)). \label{prS} 
\label{eq:ressing}
\end{eqnarray}
The pole-residue expansion controls the behavior of the solutions of
Eq.~(\ref{Keq}) at low $x$.  Since the subsequent residues carry
powers of $A^n=(Q b)^n$, the expansion cannot be used for $Q b$ too
large.

\end{document}